\documentclass[aps,prl,reprint,showpacs]{revtex4-1}
\usepackage[utf8]{inputenc}
\usepackage[T1]{fontenc}
\usepackage{amsmath, amssymb, amsfonts, graphicx, braket, color, units,lmodern}
\usepackage[colorlinks=true, linkcolor=blue, citecolor=blue, urlcolor=blue]{hyperref}

\newcommand{\me}[1]{\langle\colon{#1}\colon\rangle}

\begin{document}

\title{Single-Mode Parametric-Down-Conversion States with 50 Photons as a Source for Mesoscopic Quantum Optics}

\author{Georg Harder$^{1}$}
\email{georg.harder@uni-paderborn.de}
\author{Tim J. Bartley$^{1,2}$}
\author{Adriana E. Lita$^{2}$}
\author{Sae Woo Nam$^{2}$}
\author{Thomas Gerrits$^{2}$}
\author{Christine Silberhorn$^{1}$}

\affiliation{$^{1}$Integrated Quantum Optics Group, Applied Physics, University of Paderborn, 33098 Paderborn, Germany}
\affiliation{$^{2}$National Institute of Standards and Technology, 325 Broadway, Boulder, CO 80305, USA}

\date{\today}


\begin{abstract}
We generate pulsed, two mode squeezed states in a single spatio-temporal mode with mean photon numbers up to $20$. 
We directly measure photon-number-correlations between the two modes with transition edge sensors up to 80 photons per mode. 
This corresponds roughly to a state-dimensionality of $6400$.
We achieve detection efficiencies of $64\%$ in the technologically crucial telecom regime and demonstrate the high quality of our measurements by heralded nonclassical distributions up to $50$ photons per pulse and calculated correlation functions up to $40^\mathrm{th}$ order.
\end{abstract}

\maketitle

\paragraph*{Introduction.\ --}
The quest to study quantum effects for macroscopic system sizes is driven by one of the most fundamental issues of quantum physics, as exemplified by Schrödinger’s cat states \cite{schrodinger_gegenwartige_1935}, and has initiated much research work over the last decades \cite{leggett_macroscopic_1980, leggett_testing_2002, annett_superconductivity_2004, aspelmeyer_cavity_2014}.
However, the nature of quantum decoherence renders the observation of nonclassical features in large systems increasingly difficult. Optical states are a good candidate to observe nonclassical features and to harness large systems for new quantum applications \cite{walmsley_quantum_2015}, since they only suffer from loss as decoherence mechanism and current development of low-loss equipment enables a new generation of experiment. Crucial for both applications and fundamental questions, in the optical domain, is the ability to generate large photonic states in well-defined optical modes \cite{rohde_spectral_2007} as well as detecting them with sufficient efficiency. Starting with the landmark experiment by Hanbury-Brown and Twiss \cite{hanbury-brown_correlation_1956}, the statistical properties of photons have been used in a broad range of contexts to observe and exploit non-classical effects.

Two-mode squeezed states with large photon numbers can be considered macroscopic \cite{oudot_two-mode_2015} as they exhibit a large Fisher information \cite{braunstein_statistical_1994}. Using the process of parametric down-conversion (PDC), bright squeezed states with billions of photons have been demonstrated \cite{heidmann_observation_1987, aytur_pulsed_1990, smithey_sub-shot-noise_1992, bondani_sub-shot-noise_2007, agafonov_two-color_2010,allevi_statistics_2014, sharapova_schmidt_2015}. 
However, the multi-mode nature of this approach frequently impairs the direct comparison between theoretical predictions and experimental observations and limits the applications of these states.
In particular, 
further processing with non-Gaussian measurements projects multi-mode states into mixed states, thereby diminishing significantly the quantum character. Contrariwise, the combination of photon number measurements with genuine single- or two-mode squeezed vacuum states has been shown to overcome Gaussian no-go theorems \cite{eisert_distilling_2002}, to enable continuous variable entanglement distillation  \cite{takahashi_entanglement_2010, kurochkin_distillation_2014} and to allow for the preparation of cat states \cite{ourjoumtsev_generation_2007, gerrits_generation_2010}. Recent development in transition edge sensors (TES) \cite{lita_counting_2008} and nanowire detectors \cite{marsili_detecting_2013} offers the possibility to perform photon number measurements with single photon resolution and very high efficiency. 

Tight filtering \cite{perez_bright_2014} or mode selection \cite{brecht_demonstration_2014} could be used to reduce the number of modes, at a cost of reducing the size of the systems and achievable purity due to unavoidable losses \cite{christ_theory_2014}. 
In the single photon regime pulsed PDC sources with tailored dispersion properties have been developed, which are capable to generate directly PDC states in one mode only \cite{grice_eliminating_2001, uren_generation_2005}. Such single-mode PDC states have been shown experimentally at the single photon level using bulk PDC \cite{Mosley_heralded_2008} and up to a mean photon number of $2.5$ using a waveguide \cite{eckstein_highly_2011}. When increasing the pump power further, detrimental effects might be introduced such as time ordering effects \cite{christ_theory_2014, quesada_effects_2014}, self phase modulation\cite{sundheimer_large_1993} and photorefractive damage \cite{tyminski_photorefractive_1991, wang_survey_2000}, which may degrade the mode structure. 

In this Letter, we demonstrate that an engineered PDC source remains single mode for a broad range of pump energies and allows nonclassical, non-Gaussian states of up to $50$ photons to be heralded in single-modes. Using superconducting transition edge sensors (TES), we perform photon number measurements and show nonclassical phenomena in the photon statistics with photon numbers spanning a space of dimension 80x80.  We measure the full photon-number distribution of the state, which allows us to analyze correlation functions up to 40th order, demonstrate joint photon number squeezing with unprecedented measurement resolution and show negative parity in the raw data.

\paragraph*{Source Design and Implementation. \ --}
\begin{figure}[t]
\centering
\includegraphics[width=0.48\textwidth]{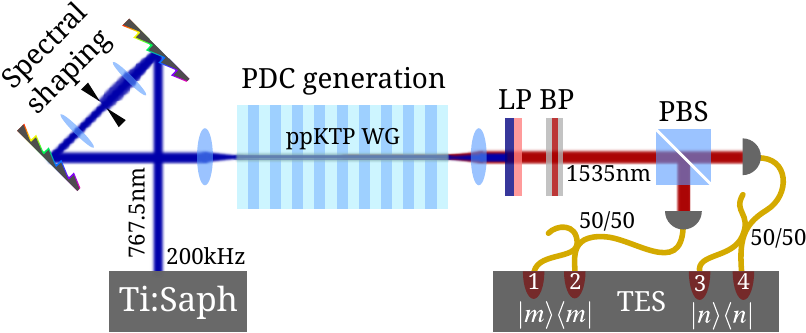}
\caption{Setup. Transform-limited pulsed light from a Ti:Sapphire laser is spectrally filtered to produce \unit[1]{ps} pulses  and coupled into the periodically poled KTP waveguide. A Long Pass (LP) filter removes the pump after the down-conversion process in the waveguide and a Band Pass (BP) filter suppresses the sinc-sidelobes of the phasematching function. Signal and idler are split at a Polarizing Beam Splitter (PBS), coupled into single-mode fibers and connected to (up to four) Transition Edge Sensors (TES).}
\label{fig:setup}
\end{figure}
We use type II PDC, where signal and idler photons are orthogonally polarized and ideally described by a product of two-mode squeezed vacuum states:
\begin{equation}
\ket{\psi} = \bigotimes_k \sqrt{1-|\lambda_k|^2} \sum_n \lambda_k^n \ket{n,n}_k,
\label{eq:PDC}
\end{equation}
where $k$ labels frequency modes, $n$ is the photon number in each mode and $\lambda = \tanh (r)$. The squeezing parameter $r$ scales linearly with the pump field amplitude, the nonlinear coefficient $\chi^{(2)}$, the interaction length inside the crystal and the mode overlap of the pump and PDC modes. Having perfect photon number correlations, the PDC states can be used to herald Fock states -- states with well-defined photon number. 

Spectrally single mode operation, i.e. $\lambda_k=0$ for $k>1$, can be achieved by engineering the momentum conservation (phasematching) condition of the nonlinear interaction \cite{grice_eliminating_2001}, which typically means engineering the nonlinear dielectric medium and pump properties. 
Spatial correlations can be fully suppressed by using a waveguide, which is single-mode for the signal and idler down-converted modes. Such single-mode generation of PDC in a waveguide has been demonstrated in \cite{eckstein_highly_2011}, yet the brightness of the source has not been explored. 

Typical approaches to generate single-mode PDC states use bulk crystals. 
To generate bright states in a bulk nonlinear medium, the pump must be tightly focused to achieve a large nonlinear interaction, which, at very high pump powers, may introduce higher-order nonlinear effects. 
A waveguide geometry has the benefit of increasing the process efficiency in a single spatial mode by two orders of magnitude compared to bulk PDC sources due to confinement of the pump beam~\cite{fiorentino_spontaneous_2007}.

\begin{figure}[t]
\centering
\includegraphics[width=0.455\textwidth]{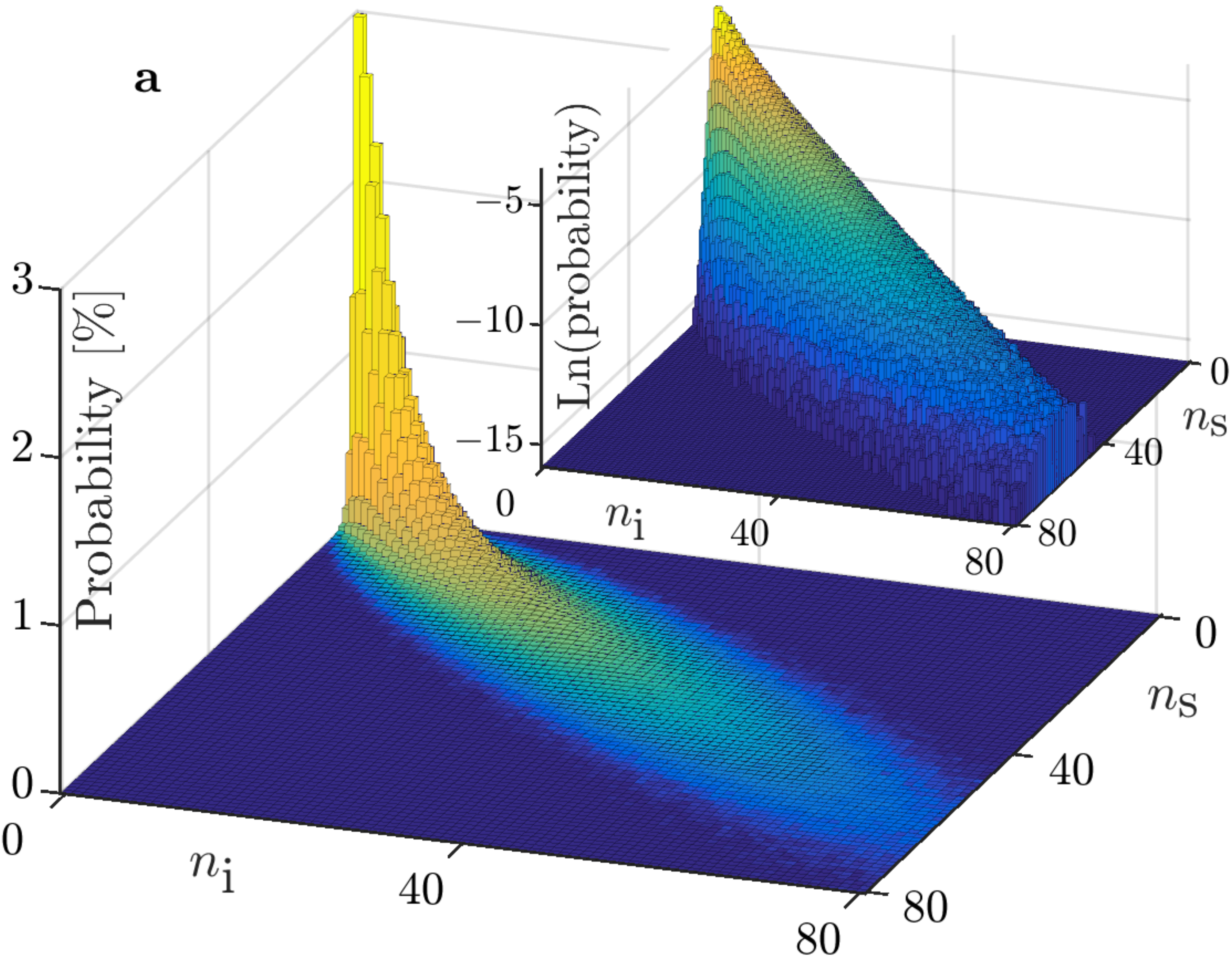}\\
\includegraphics[width=0.455\textwidth]{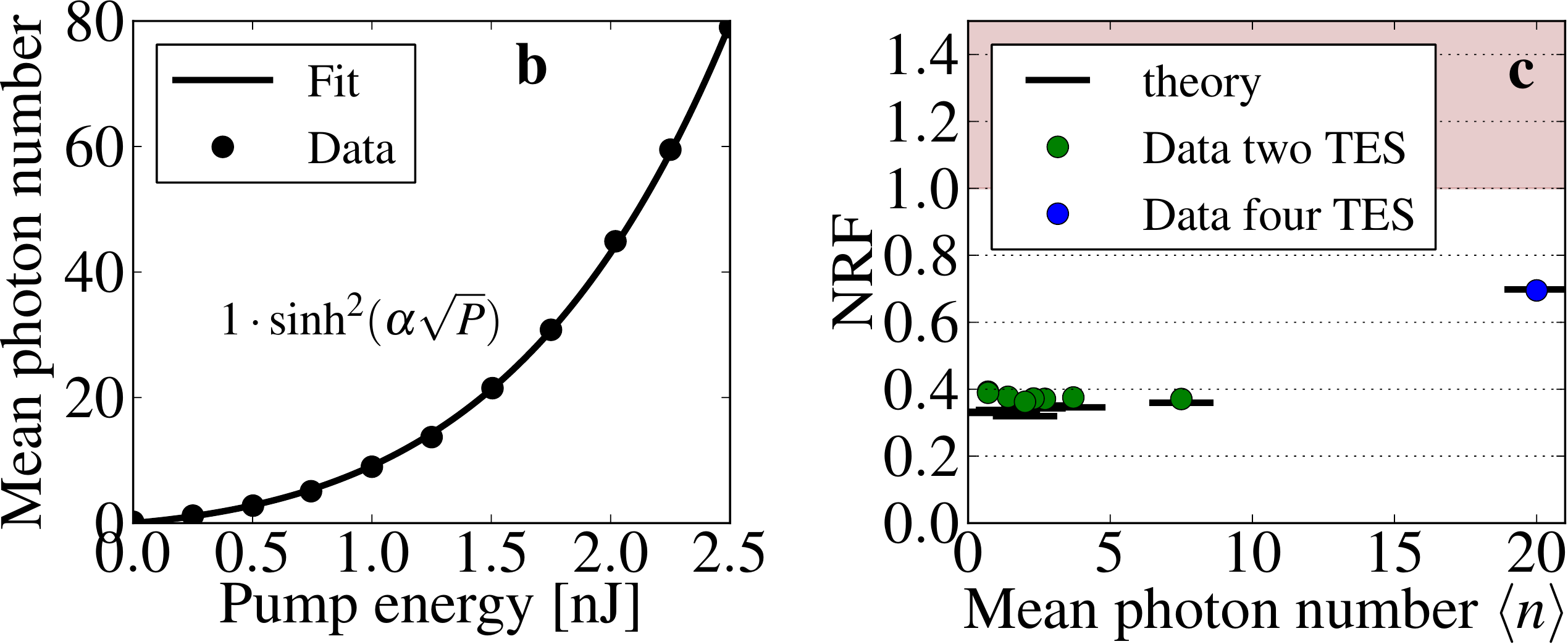}
\caption{(a) Raw photon number correlation matrix of the state $\langle n\rangle=20$ with exponentially decaying diagonal elements (inset logarithmic scale). (b) Mean photon number in one mode versus pump power measured with a low efficiency APD. The excellent fit with only one fit parameter $\alpha$ indicates that the state stays single-mode up to at least $80$ photons. (c) Noise reduction factor (NRF) for different mean photon numbers, showing the nonclassical correlations in the state. Statistical errorbars in (b) and (c) are smaller than the datapoints.}
\label{fig:shoe}
\end{figure}

A schematic of our experimental setup is shown in fig. \ref{fig:setup}. 
In the single photon regime, the ppKTP waveguide source has shown single-photon purities above $80\%$\cite{harder_optimized_2013}. 
The nonlinear medium consists of an $\unit[8]{mm}$ long periodically poled KTP waveguide engineered to produce decorrelated and degenerate signal and idler modes at $\unit[1535]{nm}$. 
The chip has been commercially purchased from ADVR.
We pump the chip with $\unit[1]{ps}$ optical pulses containing energies of up to $\unit[2.5]{nJ}$ and producing states with a mean photon number of up to $80$ photons. 
For pump pulse energies up to 1.5nJ, we measure the photon numbers, shot to shot, with transition edge sensors (TES)\cite{lita_counting_2008}. 
The TES have a near unity detection efficiency and feature single-photon resolution below 20 photons at $\unit[1535]{nm}$ but can detect up to ~100 photons with a few-photon uncertainty \cite{gerrits_extending_2012}. 
We analyze the TES response for each event based on trace overlaps with calibration traces from known coherent state inputs (see supplemental material \cite{supp} for further details). 
We use either one TES on each mode for states with mean photon numbers $\langle n \rangle < 10$ or two TES on each mode for one state with $\langle n \rangle = 20$. 
Additionally, we use an avalanche photo diode (APD) with calibrated attenuators to measure mean photon numbers up to 80.

\paragraph*{Results. \ --}
The measured photon number probabilities, shown in fig. \ref{fig:shoe}(a) for the state $\langle n\rangle =20$, feature photon number correlations as well as a logarithmic decaying diagonal, as expected from eq. \ref{eq:PDC} with only one spectral mode. 
The vacuum component is still the highest element despite measured mean photon numbers of $11$ and $9$ in each mode. 
This directly reveals the single-mode character of the state; for a multimode state, the mixture of different thermal distributions would tend towards a Poissonian distribution as the number of modes increases. 
To quantify the singlemodeness, we calculate the second order autocorrelation function \cite{loudon_quantum_2000} $g^{(2)}(0)=\frac{\langle n^2\rangle-\langle n\rangle}{\langle n\rangle^2}$, where $n$ is the photon number, on the marginal distribution of each mode. 
For thermal statistics, $g^{(2)}(0)=2$ and for Poissonian statistics $g^{(2)}(0)=1$. 
For the state shown in fig. \ref{fig:shoe}(a) we obtain $1.89(3)$ and $1.87(3)$ for signal and idler, respectively. 
This corresponds to effective mode numbers \cite{eberly_schmidt_2006, christ_probing_2011} $K=1/(g^{(2)}(0)-1)$ of $1.12(4)$ and $1.15(4)$, where $1$ would be the ideal case. 
All uncertainties given in this letter correspond to the $1\sigma$ standard deviation. We see no dependence of the effective mode number on pump power. 
When we use the highest pump powers available to us, the source generates states with a mean photon number of 80, corresponding to $r=2.9 (\lambda=0.99)$. 
The mean photon numbers as a function of pump power follow the expected curve for this measurement up to the highest available powers, see fig.\ref{fig:shoe}(b) . (For this single measurement we use an APD, calibrated using the Klyshko method\cite{klyshko_use_1980}.)

The nonclassicality of our state can be seen directly in the raw data. 
Any classical state, by definition, can be written as a mixture of coherent states with a positive probability distribution. 
Hence, the photon number uncertainty of $\sqrt{N}$ in a pulse with a mean photon number of $N$ imposes a lower bound on the antidiagonal width $n_s-n_i$ in fig. \ref{fig:shoe}(a). 
To encapsulate this criterion, one figure of merit is the noise reduction factor \cite{aytur_pulsed_1990} $\textrm{NRF}=\frac{\textrm{Var}(n_s-n_i)}{\langle n_s + n_i\rangle}$, which is necessarily $\geq1$ for classical states. For ideal PDC with a detection efficiency of $\eta$, the NRF is equal to $1-\eta$. 
We measure values below $0.4$, see fig. \ref{fig:shoe}(c), in those cases where we use one TES on each mode, in agreement with the measured efficiencies of around $66\%$. 
This corresponds to $\unit[4.2]{dB}$ of correlated photon number squeezing not corrected for losses. 
In the case where we use two TES on each mode the NRF is higher due to slightly lower and more asymmetric efficiencies in that configuration. 
\begin{figure}[t]
\centering
\includegraphics[width=0.455\textwidth]{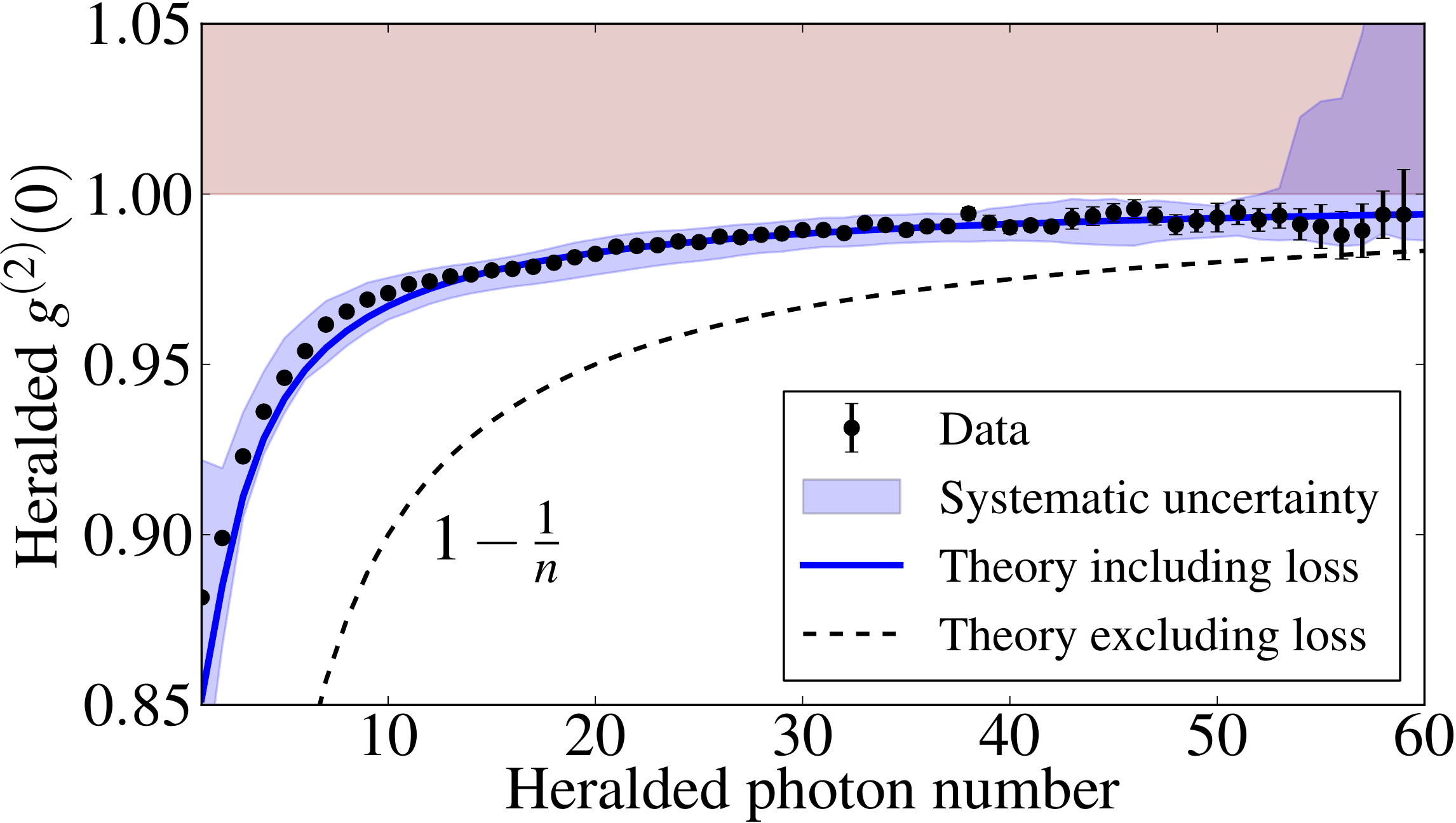}
\caption{Heralded $g^{(2)}(0)$ as a nonclassicality measure for a state with $\langle n \rangle =7$. The shaded green area accounts for worst case systematic errors stemming from the analysis of the TES response. Errorbars are statistical errors. The heralded states stay nonclassical up to around 50 photons.}
\label{fig:g2}
\end{figure}

The nonclassicality can also be seen in heralded states. For one and three-photon heralded states we see negative parities $\langle (-1)^n\rangle$ of $-0.131(1)$ and $-0.013(2)$ in the raw heralded data, which is a sufficient condition for nonclassicality. For higher heralded states the parity tends to zero and is obscured by statistical errors. 

A more robust criterion is the heralded $g^{(2)}(0)$ value, i.e. the $g^{(2)}(0)$ in one mode conditioned on a certain outcome in the other mode. For ideal $n$-photon Fock states $g^{(2)}(0)=1-1/n$. Values below 1 indicate nonclassical subpoissonian statistics. Even heralding on a 50-photon event, the measured states fulfill this nonclassicality criterion, see fig. \ref{fig:g2}. With increasing photon  number, the transition from strongly nonclassical states to classical states becomes apparent as they become harder to distinguish. Producing larger nonclassical states would require reducing the losses in the heralding mode. At the current efficiencies, the $50$ photon event happens about twice per second with a PDC mean photon number of $7$. 

\begin{figure}[t]
\centering
\includegraphics[width=0.455\textwidth]{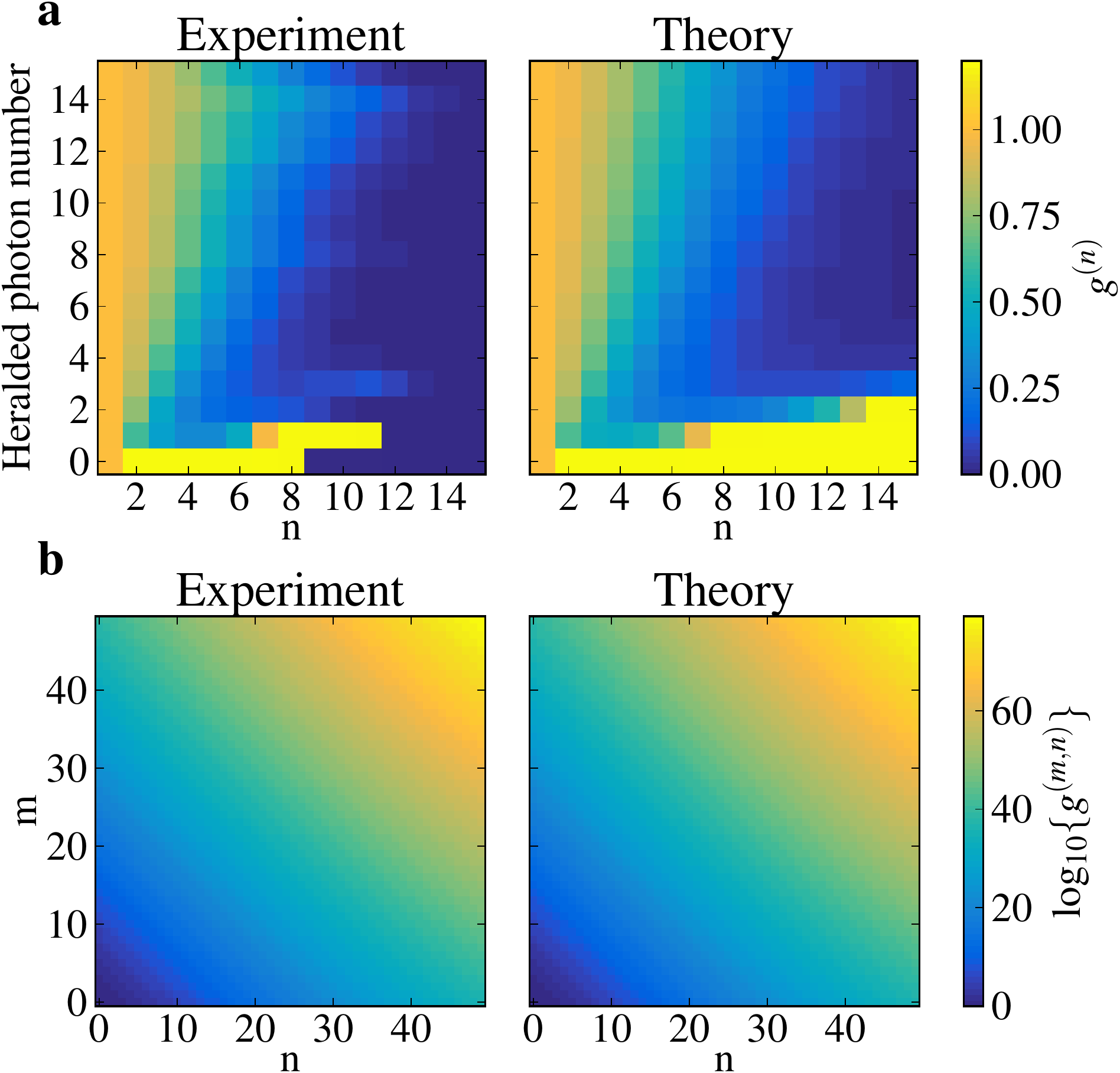}
\caption{Correlation functions. (a) $g^{(n)}$ for heralded states from a PDC state with $\langle n\rangle=1.4$. Experimental results are shown on the left and theoretical predictions on the right. Note that values smaller than one imply nonclassicality. (b) $g^{(m,n)}$ for a PDC state with $\langle n \rangle = 20$. See supplemental material for further details \cite{supp}.}
\label{fig:gmn}
\end{figure}

Having access to the full photon number distribution allows us to go beyond the well-established second order $g^{(2)}(0)$ and calculate higher order correlation function which may be defined by $g^{(n)} = \frac{\langle a^{\dagger n} a^n \rangle}{\langle a^\dagger a\rangle^n} $for one mode and $ g^{(m,n)} = \frac{\langle a^{\dagger m} a^m b^{\dagger n} b^n\rangle}{\langle a^\dagger a\rangle^m\langle b^\dagger b\rangle^n}$ for two modes, where $a$, $a^\dagger$ and $b$, $b^\dagger$ are the usual annihilation and creation operators.
In figure \ref{fig:gmn} we show the results of both cases. The measured values are in excellent agreement with the theoretical predictions. Further details on correlation functions are given in the supplemental material \cite{supp}.

The excellent agreement with theory indicates that the limiting factor is indeed the loss in our setup. We calculate our system efficiencies by either assuming perfect photon number correlations \cite{worsley_absolute_2009} or by a least square fit (see supplementary material \cite{supp}). We obtain $60\%$ and  $64\%$ for signal and idler, respectively, using the first method and $64\%$ and $68\%$ using the second method, with systematic uncertainties around $3\%$. The efficiencies are slightly higher in the latter case because we allow for Poissonian and thermal noise in the original data stemming either from an optical background or a non-perfect photon number resolution in the detectors. Such noise behaves like loss in the first method. For the $\langle n\rangle = 20$ state, the second method gives $43\%$ and $52\%$ for signal and idler. Here, the efficiencies are lower due to the change of the experimental configuration from two TES to four TES requiring an extra pair of fibre beam splitters. 

Total efficiencies close to $70\%$ are among the highest in the literature \cite{giustina_bell_2013, christensen_detection-loophole-free_2013, ramelow_highly_2013, dixon_heralding_2014}. These high efficiencies are the reason why we see clear nonclassical features in the raw data without loss inversion. For example, negative parity can only be observed above $50\%$ in principle. 

Given the performance of the source, we can estimate the potential continuous variable (CV) squeezing \cite{loudon_quantum_2000} achievable with the current setup. In the literature, to our knowledge, the highest squeezing directly measured in a single pass, pulsed system is $\unit[5]{db}$ \cite{eto_efficient_2011} and in a continuous-wave cavity system $\unit[12.7]{dB}$ \cite{eberle_quantum_2010}. 
In our source, the maximum mean photon number of $80$ would correspond to $\unit[25]{dB}$ of squeezing. The measurable squeezing, however, would be limited by the current efficiencies to about $\unit[4.5]{dB}$.  

The main loss contributions in the setup come from the coupling to single-mode fibers of around $80\%$ and the linear optical elements with a total transmission of about $90\%$. With on chip integration of polarizing beam splitters and detectors, of which both have been demonstrated \cite{alferness_low-cross-talk_1984, calkins_high_2013}, the total efficiencies could go up to above $90\%$. This would push the size of possible nonclassical states to hundreds of photons. The ultimate goal would be an efficiency around $99\%$ at which fault tolerant quantum computation with CV cluster states becomes possible \cite{menicucci_fault-tolerant_2014}.

\paragraph*{Conclusion. \ --}

Observing nonclassical correlations of photons is fundamental to quantum optics. We have shown that these correlations persist with the largest number of photons to date in a single-mode state. 
The single-mode nature of these states allow us to herald large photon-number states in a controlled and efficient way. When combined with non-Gaussian projective measurements and homodyne detection, a broad range of continuous-variable experiments in the strongly-squeezed, pulsed regime become possible.

\paragraph*{Acknowledgment. \ --}
This work was supported by the Quantum Information Science Initiative (QISI) and the Deutsche Forschungsgemeinschaft (DFG) via TRR142 and via the Gottfried Wilhelm Leibniz-Preis. TJB acknowledges funding from the Deutscher Akademischer Austausch Dienst (DAAD), and LK Shalm for discussions.

Contribution of NIST, an agency of the US government, not subject to copyright

G.H. and T.J.B. contributed equally to this work.


\newpage

\begin{center}
\textbf{\Large{Supplemental Material}}
\end{center}


\section{Correlation Functions}
Correlation functions, first introduced by Glauber \cite{glauber_quantum_1963},  are one way to characterize photon number statistics. For example, varieties of nonclassicality criteria can be constructed based on second and higher order moments of the electromagnetic field \cite{vogel_nonclassical_2008,miranowicz_testing_2010}. Such criteria identify nonclassical fields directly from the measured statistics in a loss tolerant way without complicated analysis techniques and have been utilized with low order correlation functions \cite{avenhaus_accessing_2010,sperling_uncovering_2015}. Having access to higher order correlation functions gives a more complete description of the underlying photon number statistics and increases the possibilities in characterizing quantum states. For example, exotic quantum states might only reveal their nonclassicality when higher order correlations are included. In this section we show that we are able to calculate these higher order correlation functions.

We focus on correlation functions of the form  $g^{(n)} = \frac{\langle a^{\dagger n} a^n \rangle}{\langle a^\dagger a\rangle^n} $for one mode or $g^{(m,n)} = \frac{\langle a^{\dagger m} a^m b^{\dagger n} b^n\rangle}{\langle a^\dagger a\rangle^m\langle b^\dagger b\rangle^n}$ for two modes, where $a$, $a^\dagger$ and $b$, $b^\dagger$ are the usual annihilation and creation operators. They can be calculated from photon number probabilities $p_k$ using $\langle a^{\dagger n} a^n \rangle = \sum_k \prod_{l=0}^{n-1} (k-l) p_k$. 

\begin{figure}[ht]
\centering
\includegraphics[width=0.48\textwidth]{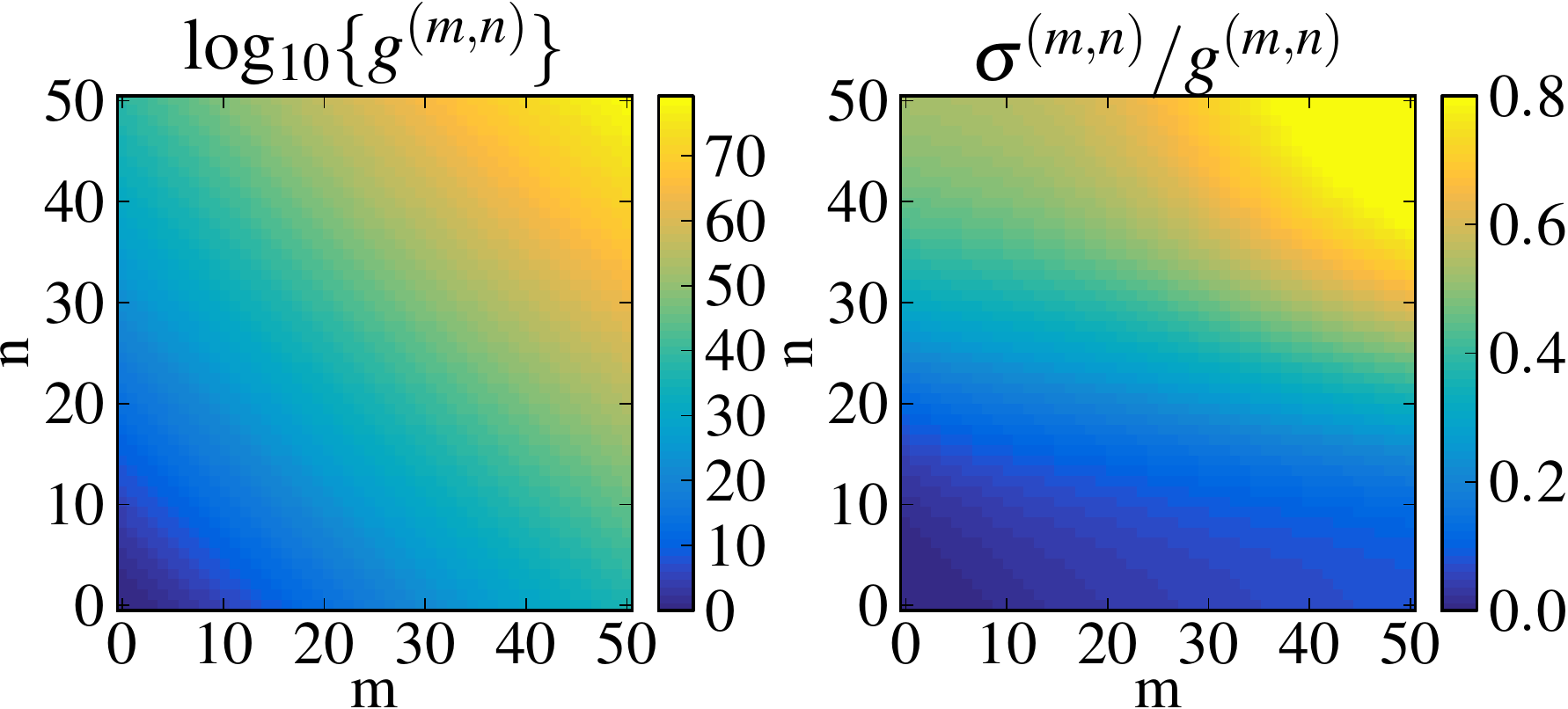}
\caption{Left: $g^{(m,n)}$ for the state with $\langle n\rangle=20$. Right: Relative error obtained from a Monte-Carlo simulation based on the measured probability distribution. Values up to $g^{(40,40)}$ seem reliable. The asymmetry in the two modes arises from asymmetric detection efficiencies.}
\label{fig:sup_gmn}
\end{figure}
\begin{figure}[ht]
\centering
\includegraphics[width=0.48\textwidth]{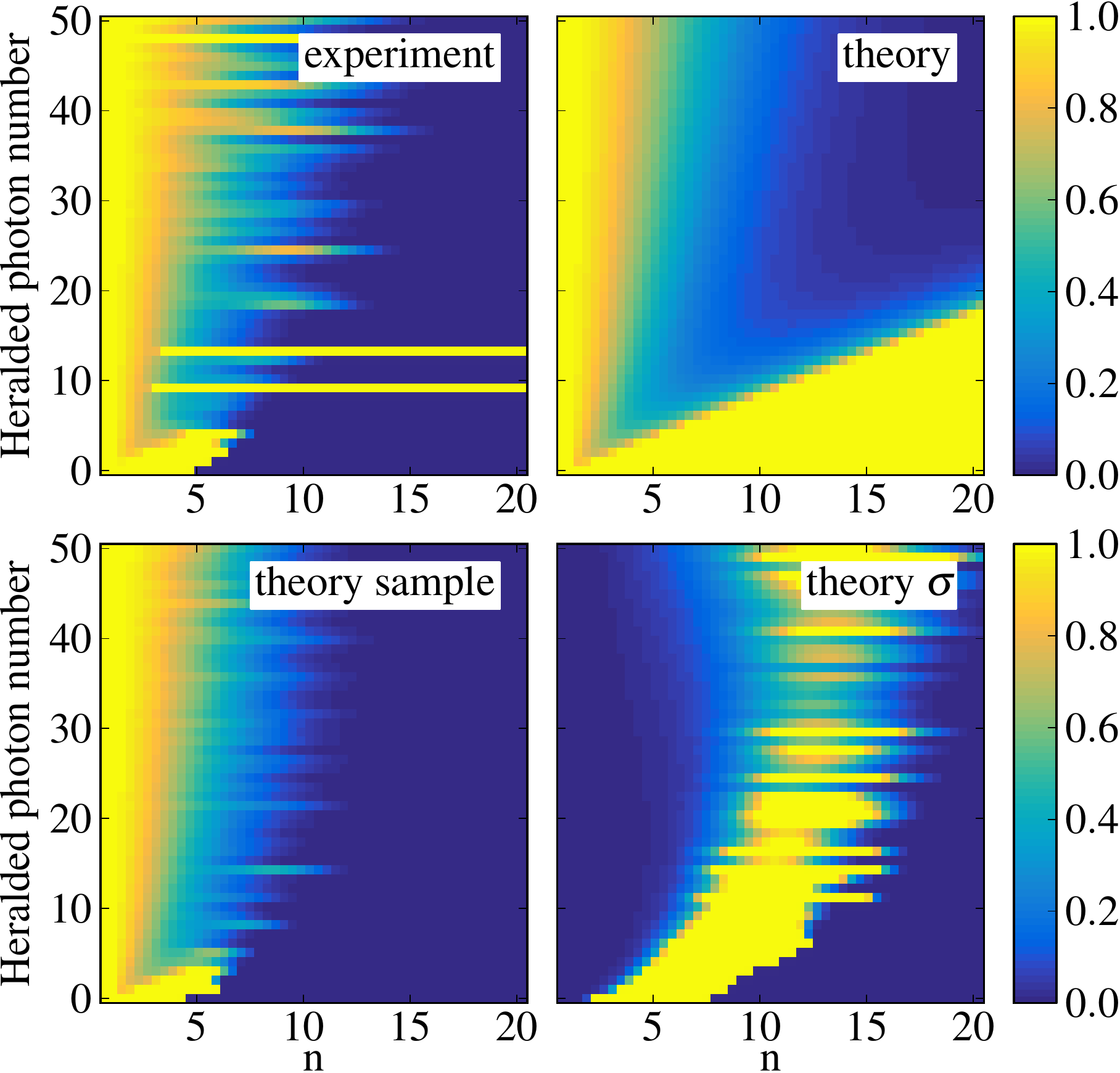}
\caption{Correlation functions $g^{(n)}$ for heralded states from a PDC state with $\langle n\rangle=7$. Top: Comparison of experiment and theory. Bottom left: Simulation with $8.2 \cdot 10^6$ events. Bottom right: Expected standard deviation from a Monte-Carlo simulation. Only the region left of the yellow area is reliable with the given statistics.}
\label{fig:sup_gn}
\end{figure}

In fig. \ref{fig:sup_gmn} we show the two mode $g^{(m,n)}$ for the bright state $\langle n\rangle=20$. We estimate the uncertainties of the values by performing a simple Monte-Carlo simulation: Based on the measured photon  number probabilities, we draw measurement frequencies from $8.2\cdot 10^6$ events and calculate the according correlation functions. From 10000 such trials, we calculate the standard deviations. Values up to $g^{(40,40)}$ seem to have reasonably low values.

In fig. \ref{fig:sup_gn} we show $g^{(n)}$ values for heralded states from the PDC state $\langle n\rangle=7$. For the error analysis, we perform the same Monte-Carlo simulation, though this time based on theoretical probability distributions, due to relatively low numbers of heralded events for higher photon numbers. For example, the 35-photon herald happens only about 1000 times. This is one reason, why the statistical uncertainties dominate already above $g^{(10)}$. However, the general agreement with theory is very good. Almost all values are significantly below one, which is the bound for classical states.

\section{Nonclassicality}
Following \cite{miranowicz_testing_2010}, we construct the correlation matrix
\begin{equation*}
M = 
\begin{pmatrix}
\me{\hat f^\dagger_1 \hat f_1} & \me{\hat f^\dagger_1 \hat f_2} & \cdots & \me{\hat f^\dagger_1 \hat f_N} \\
\me{\hat f^\dagger_2 \hat f_1} & \me{\hat f^\dagger_2 \hat f_2} & \cdots & \me{\hat f^\dagger_2 \hat f_N} \\
\vdots & \vdots & \ddots & \vdots \\
\me{\hat f^\dagger_N \hat f_1} & \me{\hat f^\dagger_N \hat f_2} & \cdots & \me{\hat f^\dagger_N \hat f_N} \\
\end{pmatrix},
\end{equation*}
where $\vec{\hat f} = (1, \hat n_a, \hat n_b, \hat n_a^2, \hat n_a \hat n_b, \hat n_b^2, ...,\hat n_b^N)$, $\hat n_a=a^\dagger a/2$, $\hat n_b = b^\dagger b/2$ and $\colon \colon$ denotes normal ordering.
If $M$ has at least one negative eigenvalue, the state is nonclassical. This condition is fulfilled for all states presented here. To estimate the uncertainties, we again apply a Monte-Carlo simulation. In the best case, for the $\langle n\rangle=7$ state, the lowest eigenvalue has a significance of more than 100 standard deviations. This shows the high quality of the measured statistics.

\section{Loss Inversion}
To get a glimpse of how our states would look without losses, we fit a model to the data. The model consists of a state that can be described as a mixture of a (spectrally) multimode PDC state, a coherent state and a thermal state:
\begin{equation}
\rho = \rho_\textrm{PDC}(n^\textrm{PDC}, K) \otimes  \rho_\alpha(n^{\alpha}_\textrm{s}, n^{\alpha}_\textrm{i}) \otimes  \rho_\textrm{th}(n^\textrm{th}_\textrm{s}, n^\textrm{th}_\textrm{i}),
\label{eq:sup_model}
\end{equation} 
where $n$ are the respective mean photon numbers and $K$ the effective mode number of the PDC state. We expect $K$ to be low since the marginal $g^{(2)}(0)$ measurements mentioned in the paper suggest $K\approx 1.13$. We hence choose exponentially decaying coefficients $\lambda_k^2$ for each (spectral) mode, whereas $ \sum_k\lambda_k^2=1$ and $K=1/\sum_k \lambda_k^4$ \cite{eberly_schmidt_2006, christ_probing_2011}. The squeezing parameters for each (spectral) PDC mode are given by $r_k = B\lambda_k$, where $B$ is the overall optical gain. Such exponentially decaying coefficients are a reasonable approximation for low effective mode numbers.

The losses are described by a standard beam splitter model with transmissions $\eta_s$ and $\eta_i$ in the two beam paths. The photon number probabilities are given by
$p^\textrm{out}_{kl} = \sum_{mn}L_{km}^\textrm{s}(\eta_\textrm{s}) L_{ln}^\textrm{i}(\eta_\textrm{i}) p^\textrm{in}_{mn},$ and $L_{kn}(\eta) = \binom{n}{k} \eta^k(1-\eta)^{n-k}$.
In eq. \ref{eq:sup_model}, the photon number distributions of the three contributions are independent. That means that the total photon number distribution $p^\textrm{in}$ is a convolution of the three individual distribution. This can be implemented numerically in a straight forward way. 

Finally, we minimize the weighted sum of the  least square differences 
\begin{equation}
{\sum_{mn}((p^\mathrm{meas}_{mn}-p^\textrm{out}_{mn})/\sigma_{mn})^2},
\end{equation}
where $p^\mathrm{meas}_{mn}$ are the measured photon number probabilities, $p^\textrm{out}_{mn}$ the probabilities of the expected state after losses and $\sigma_{mn}=1/N + \sqrt{p^\mathrm{meas}_{mn}/N}$ estimates for the statistical error due to $N$ total events. Effectively, this is a fit with eight parameters $(\eta_s, \eta_i, n_\textrm{PDC}, K, n^{\alpha}_\textrm{s},  n^{\alpha}_\textrm{i}, n^\textrm{th}_\textrm{s}, n^\textrm{th}_\textrm{i})$. Allowing Poissonian and thermal background statistics covers most optical and electrical background signals while keeping the number of free parameters very low. 

\begin{figure}[t]
\centering
\includegraphics[width=0.455\textwidth]{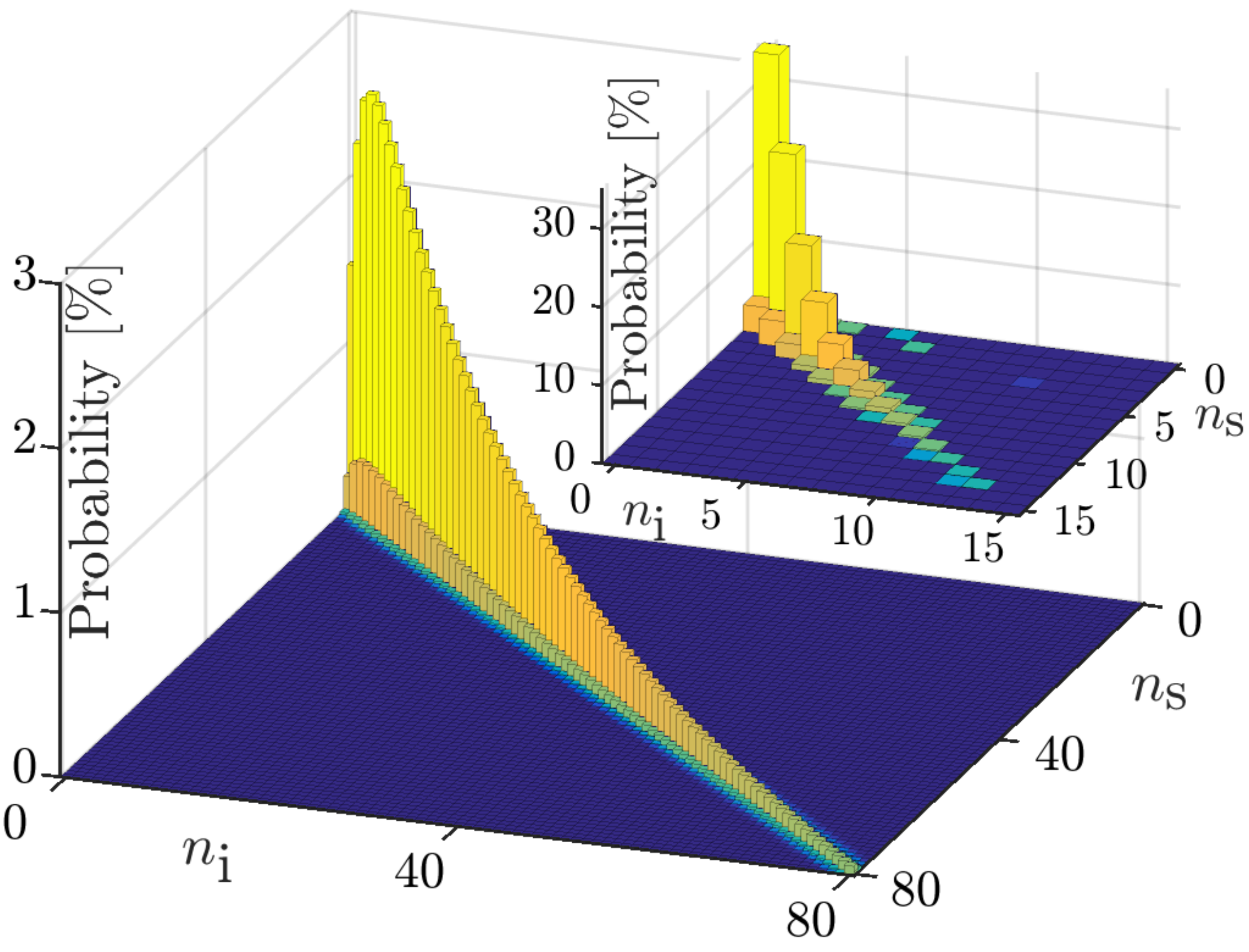}
\caption{Inferred states before losses. a) High power state ($\langle n\rangle = 20$) using a parameter fit to the data. b) Low power state ($\langle n\rangle = 1.4$) using full loss inversion without assumptions about the state.}
\label{fig:inversion}
\end{figure}

The fit result for the state with $\langle n \rangle = 20$ is shown in fig. \ref{fig:inversion} and has the fit parameters 
\begin{center}
\begin{tabular}{|c|c|}
\hline
$\eta_\textrm{s}$ & $43.13(3)\%$ \\
$\eta_\textrm{i}$  & $52.12(4)\%$ \\
$n^\textrm{PDC}$ & $20.30(2)$ \\
$K$ & $1.097(1)$ \\
$n^{\alpha}_\textrm{s}$ & $0.14(12)$ \\
$n^{\alpha}_\textrm{i}$ &  $0.38(5)$ \\
$n^\textrm{th}_\textrm{s}$ & $0.00(12)$  \\
$n^\textrm{th}_\textrm{i}$ & $0.00(5)$ \\
\hline
\end{tabular}
\end{center}

The fidelity with the data is $99.98\%$. The largest contribution in photon number by almost two orders of magnitude is the PDC. Furthermore, the effective mode number is low in agreement with the $g^{(2)}(0)$ measurements. 

We can also regard such a fit as an efficiency estimation that is not impacted by non-PDC counts. In the case of our more efficient two-TES setup configuration with the $\langle n \rangle = 7$ state, it suggests efficiencies of $64\%$ and $68\%$. 

For comparison, we perform a general loss inversion\cite{feito_measuring_2009} for a low power state with $\langle n \rangle = 1.4$, shown in fig. \ref{fig:inversion}(inset), restricting the space to $<15$ photons. Again, the inverted state resembles the expected PDC state very well. The number of free parameters is very high ($15^2-1$ in this example) such that general loss inversion becomes infeasible for states with higher mean photon numbers.

\section{Methods: Trace Analysis}
Each detection event of the TES is a voltage pulse $V(t)$ whose shape depends on the detected photon number. The characteristics of these traces are different for each TES and the photon numbers cannot be well distinguished by peak height alone. However, good photon number resolution can be obtained by simply using the average trace as a template $\bar V(t)$ and take the overlap $\int \textrm{d} t V(t)\bar V(t)$ to distinguish between photon numbers\cite{levine_algorithm_2012}. This method is ideal if the shapes of all pulses are similar and works well for coherent input light. In our case, however, the full range of photon numbers is present and the shapes are different for high and low photon numbers. We therefore adapt this technique as follows: We calibrate the TES responses using coherent input light. For $20$ different input power settings, we calculate the average traces $\bar V_i(t)$ and calibrate the overlaps of each based on the Poissonian photon number expectations. Since Poissonian distributions are relatively narrow, this calibration is only reliable around the respective mean photon numbers. Then, for an unknown detection event, we calculate the overlaps with $20$ templates giving $20$ photon number estimations, ideally all the same, and take that particular estimation that is closest to the mean photon number of its template. This method extends the range over which we can reliably resolve photon number as compared to the one-template approach and even gives reliable estimations of photon numbers beyond the single-photon resolution regime. The clustering of the overlaps can still be seen up to $20$ photons in a histogram, allowing for cross checking the calibration simply by counting peaks. To estimate systematic uncertainties in the $g^{(2)}(0)$ results of fig. 3, we rescale all templates slightly until the photon numbers are clearly over- or underestimated  giving a worst case (or maximum range) effect of our photon number estimate on $g^{(2)}(0)$.

\bibliographystyle{apsrev4-1}


\begin{thebibliography}{52}%
\makeatletter
\providecommand \@ifxundefined [1]{%
 \@ifx{#1\undefined}
}%
\providecommand \@ifnum [1]{%
 \ifnum #1\expandafter \@firstoftwo
 \else \expandafter \@secondoftwo
 \fi
}%
\providecommand \@ifx [1]{%
 \ifx #1\expandafter \@firstoftwo
 \else \expandafter \@secondoftwo
 \fi
}%
\providecommand \natexlab [1]{#1}%
\providecommand \enquote  [1]{``#1''}%
\providecommand \bibnamefont  [1]{#1}%
\providecommand \bibfnamefont [1]{#1}%
\providecommand \citenamefont [1]{#1}%
\providecommand \href@noop [0]{\@secondoftwo}%
\providecommand \href [0]{\begingroup \@sanitize@url \@href}%
\providecommand \@href[1]{\@@startlink{#1}\@@href}%
\providecommand \@@href[1]{\endgroup#1\@@endlink}%
\providecommand \@sanitize@url [0]{\catcode `\\12\catcode `\$12\catcode
  `\&12\catcode `\#12\catcode `\^12\catcode `\_12\catcode `\%12\relax}%
\providecommand \@@startlink[1]{}%
\providecommand \@@endlink[0]{}%
\providecommand \url  [0]{\begingroup\@sanitize@url \@url }%
\providecommand \@url [1]{\endgroup\@href {#1}{\urlprefix }}%
\providecommand \urlprefix  [0]{URL }%
\providecommand \Eprint [0]{\href }%
\providecommand \doibase [0]{http://dx.doi.org/}%
\providecommand \selectlanguage [0]{\@gobble}%
\providecommand \bibinfo  [0]{\@secondoftwo}%
\providecommand \bibfield  [0]{\@secondoftwo}%
\providecommand \translation [1]{[#1]}%
\providecommand \BibitemOpen [0]{}%
\providecommand \bibitemStop [0]{}%
\providecommand \bibitemNoStop [0]{.\EOS\space}%
\providecommand \EOS [0]{\spacefactor3000\relax}%
\providecommand \BibitemShut  [1]{\csname bibitem#1\endcsname}%
%
\@ifxundefined \urlstyle {%
  \providecommand \doi [1]{doi:\discretionary{}{}{}#1}%
}{%
  \providecommand \doi [0]{doi:\discretionary{}{}{}\begingroup
  \urlstyle{rm}\Url }%
}%
\providecommand \doibase [0]{http://dx.doi.org/}%
\providecommand \Doi[1]{\href{\doibase#1}}%
\providecommand \bibAnnote [3]{%
  \BibitemShut{#1}%
  \begin{quotation}\noindent
    \textsc{Key:}\ #2\\\textsc{Annotation:}\ #3%
  \end{quotation}%
}%
\providecommand \bibAnnoteFile [2]{%
  \IfFileExists{#2}{\bibAnnote {#1} {#2} {\input{#2}}}{}%
}
%
\let\auto@bib@innerbib\@empty
\bibitem [{\citenamefont {Schrödinger}(1935)}]{schrodinger_gegenwartige_1935}%
  \BibitemOpen
  \bibfield  {author} {\bibinfo {author} {\bibfnamefont {E.}~\bibnamefont
  {Schrödinger}},\ }\href {\doibase 10.1007/BF01491891} {\bibfield  {journal}
  {\bibinfo  {journal} {Naturwissenschaften}\ }\textbf {\bibinfo {volume}
  {23}},\ \bibinfo {pages} {807} (\bibinfo {year} {1935})}\BibitemShut
  {NoStop}%
\bibitem [{\citenamefont {Leggett}(1980)}]{leggett_macroscopic_1980}%
  \BibitemOpen
  \bibfield  {author} {\bibinfo {author} {\bibfnamefont {A.~J.}\ \bibnamefont
  {Leggett}},\ }\href {\doibase 10.1143/PTPS.69.80} {\bibfield  {journal}
  {\bibinfo  {journal} {Prog. Theor. Phys. Supplement}\ }\textbf {\bibinfo
  {volume} {69}},\ \bibinfo {pages} {80} (\bibinfo {year} {1980})}\BibitemShut
  {NoStop}%
\bibitem [{\citenamefont {Leggett}(2002)}]{leggett_testing_2002}%
  \BibitemOpen
  \bibfield  {author} {\bibinfo {author} {\bibfnamefont {A.~J.}\ \bibnamefont
  {Leggett}},\ }\href {\doibase 10.1088/0953-8984/14/15/201} {\bibfield
  {journal} {\bibinfo  {journal} {J. Phys.: Condens. Matter}\ }\textbf
  {\bibinfo {volume} {14}},\ \bibinfo {pages} {R415} (\bibinfo {year}
  {2002})}\BibitemShut {NoStop}%
\bibitem [{\citenamefont {Annett}(2004)}]{annett_superconductivity_2004}%
  \BibitemOpen
  \bibfield  {author} {\bibinfo {author} {\bibfnamefont {J.~F.}\ \bibnamefont
  {Annett}},\ }\href@noop {} {\emph {\bibinfo {title} {Superconductivity,
  {Superfluids} and {Condensates}}}}\ (\bibinfo  {publisher} {OUP Oxford},\
  \bibinfo {year} {2004})\BibitemShut {NoStop}%
\bibitem [{\citenamefont {Aspelmeyer}\ \emph {et~al.}(2014)\citenamefont
  {Aspelmeyer}, \citenamefont {Kippenberg},\ and\ \citenamefont
  {Marquardt}}]{aspelmeyer_cavity_2014}%
  \BibitemOpen
  \bibfield  {author} {\bibinfo {author} {\bibfnamefont {M.}~\bibnamefont
  {Aspelmeyer}}, \bibinfo {author} {\bibfnamefont {T.~J.}\ \bibnamefont
  {Kippenberg}}, \ and\ \bibinfo {author} {\bibfnamefont {F.}~\bibnamefont
  {Marquardt}},\ }\href {\doibase 10.1103/RevModPhys.86.1391} {\bibfield
  {journal} {\bibinfo  {journal} {Rev. Mod. Phys.}\ }\textbf {\bibinfo {volume}
  {86}},\ \bibinfo {pages} {1391} (\bibinfo {year} {2014})}\BibitemShut
  {NoStop}%
\bibitem [{\citenamefont {Walmsley}(2015)}]{walmsley_quantum_2015}%
  \BibitemOpen
  \bibfield  {author} {\bibinfo {author} {\bibfnamefont {I.~A.}\ \bibnamefont
  {Walmsley}},\ }\href {\doibase 10.1126/science.aab0097} {\bibfield  {journal}
  {\bibinfo  {journal} {Science}\ }\textbf {\bibinfo {volume} {348}},\ \bibinfo
  {pages} {525} (\bibinfo {year} {2015})}\BibitemShut {NoStop}%
\bibitem [{\citenamefont {Rohde}\ \emph {et~al.}(2007)\citenamefont {Rohde},
  \citenamefont {Mauerer},\ and\ \citenamefont
  {Silberhorn}}]{rohde_spectral_2007}%
  \BibitemOpen
  \bibfield  {author} {\bibinfo {author} {\bibfnamefont {P.~P.}\ \bibnamefont
  {Rohde}}, \bibinfo {author} {\bibfnamefont {W.}~\bibnamefont {Mauerer}}, \
  and\ \bibinfo {author} {\bibfnamefont {C.}~\bibnamefont {Silberhorn}},\
  }\href {\doibase 10.1088/1367-2630/9/4/091} {\bibfield  {journal} {\bibinfo
  {journal} {New J. Phys.}\ }\textbf {\bibinfo {volume} {9}},\ \bibinfo {pages}
  {91} (\bibinfo {year} {2007})}\BibitemShut {NoStop}%
\bibitem [{\citenamefont {Hanbury-Brown}\ and\ \citenamefont
  {Twiss}(1956)}]{hanbury-brown_correlation_1956}%
  \BibitemOpen
  \bibfield  {author} {\bibinfo {author} {\bibfnamefont {R.}~\bibnamefont
  {Hanbury-Brown}}\ and\ \bibinfo {author} {\bibfnamefont {R.}~\bibnamefont
  {Twiss}},\ }\href@noop {} {\bibfield  {journal} {\bibinfo  {journal}
  {Nature}\ }\textbf {\bibinfo {volume} {177}},\ \bibinfo {pages} {27}
  (\bibinfo {year} {1956})}\BibitemShut {NoStop}%
\bibitem [{\citenamefont {Oudot}\ \emph {et~al.}(2015)\citenamefont {Oudot},
  \citenamefont {Sekatski}, \citenamefont {Fröwis}, \citenamefont {Gisin},\
  and\ \citenamefont {Sangouard}}]{oudot_two-mode_2015}%
  \BibitemOpen
  \bibfield  {author} {\bibinfo {author} {\bibfnamefont {E.}~\bibnamefont
  {Oudot}}, \bibinfo {author} {\bibfnamefont {P.}~\bibnamefont {Sekatski}},
  \bibinfo {author} {\bibfnamefont {F.}~\bibnamefont {Fröwis}}, \bibinfo
  {author} {\bibfnamefont {N.}~\bibnamefont {Gisin}}, \ and\ \bibinfo {author}
  {\bibfnamefont {N.}~\bibnamefont {Sangouard}},\ }\href {\doibase
  10.1364/JOSAB.32.002190} {\bibfield  {journal} {\bibinfo  {journal} {Journal
  of the Optical Society of America B}\ }\textbf {\bibinfo {volume} {32}},\
  \bibinfo {pages} {2190} (\bibinfo {year} {2015})}\BibitemShut {NoStop}%
\bibitem [{\citenamefont {Braunstein}\ and\ \citenamefont
  {Caves}(1994)}]{braunstein_statistical_1994}%
  \BibitemOpen
  \bibfield  {author} {\bibinfo {author} {\bibfnamefont {S.~L.}\ \bibnamefont
  {Braunstein}}\ and\ \bibinfo {author} {\bibfnamefont {C.~M.}\ \bibnamefont
  {Caves}},\ }\href {\doibase 10.1103/PhysRevLett.72.3439} {\bibfield
  {journal} {\bibinfo  {journal} {Phys. Rev. Lett.}\ }\textbf {\bibinfo
  {volume} {72}},\ \bibinfo {pages} {3439} (\bibinfo {year}
  {1994})}\BibitemShut {NoStop}%
\bibitem [{\citenamefont {Heidmann}\ \emph {et~al.}(1987)\citenamefont
  {Heidmann}, \citenamefont {Horowicz}, \citenamefont {Reynaud}, \citenamefont
  {Giacobino}, \citenamefont {Fabre},\ and\ \citenamefont
  {Camy}}]{heidmann_observation_1987}%
  \BibitemOpen
  \bibfield  {author} {\bibinfo {author} {\bibfnamefont {A.}~\bibnamefont
  {Heidmann}}, \bibinfo {author} {\bibfnamefont {R.~J.}\ \bibnamefont
  {Horowicz}}, \bibinfo {author} {\bibfnamefont {S.}~\bibnamefont {Reynaud}},
  \bibinfo {author} {\bibfnamefont {E.}~\bibnamefont {Giacobino}}, \bibinfo
  {author} {\bibfnamefont {C.}~\bibnamefont {Fabre}}, \ and\ \bibinfo {author}
  {\bibfnamefont {G.}~\bibnamefont {Camy}},\ }\href {\doibase
  10.1103/PhysRevLett.59.2555} {\bibfield  {journal} {\bibinfo  {journal}
  {Phys. Rev. Lett.}\ }\textbf {\bibinfo {volume} {59}},\ \bibinfo {pages}
  {2555} (\bibinfo {year} {1987})}\BibitemShut {NoStop}%
\bibitem [{\citenamefont {Aytür}\ and\ \citenamefont
  {Kumar}(1990)}]{aytur_pulsed_1990}%
  \BibitemOpen
  \bibfield  {author} {\bibinfo {author} {\bibfnamefont {O.}~\bibnamefont
  {Aytür}}\ and\ \bibinfo {author} {\bibfnamefont {P.}~\bibnamefont {Kumar}},\
  }\href {\doibase 10.1103/PhysRevLett.65.1551} {\bibfield  {journal} {\bibinfo
   {journal} {Phys. Rev. Lett.}\ }\textbf {\bibinfo {volume} {65}},\ \bibinfo
  {pages} {1551} (\bibinfo {year} {1990})}\BibitemShut {NoStop}%
\bibitem [{\citenamefont {Smithey}\ \emph {et~al.}(1992)\citenamefont
  {Smithey}, \citenamefont {Beck}, \citenamefont {Belsley},\ and\ \citenamefont
  {Raymer}}]{smithey_sub-shot-noise_1992}%
  \BibitemOpen
  \bibfield  {author} {\bibinfo {author} {\bibfnamefont {D.~T.}\ \bibnamefont
  {Smithey}}, \bibinfo {author} {\bibfnamefont {M.}~\bibnamefont {Beck}},
  \bibinfo {author} {\bibfnamefont {M.}~\bibnamefont {Belsley}}, \ and\
  \bibinfo {author} {\bibfnamefont {M.~G.}\ \bibnamefont {Raymer}},\ }\href
  {\doibase 10.1103/PhysRevLett.69.2650} {\bibfield  {journal} {\bibinfo
  {journal} {Phys. Rev. Lett.}\ }\textbf {\bibinfo {volume} {69}},\ \bibinfo
  {pages} {2650} (\bibinfo {year} {1992})}\BibitemShut {NoStop}%
\bibitem [{\citenamefont {Bondani}\ \emph {et~al.}(2007)\citenamefont
  {Bondani}, \citenamefont {Allevi}, \citenamefont {Zambra}, \citenamefont
  {Paris},\ and\ \citenamefont {Andreoni}}]{bondani_sub-shot-noise_2007}%
  \BibitemOpen
  \bibfield  {author} {\bibinfo {author} {\bibfnamefont {M.}~\bibnamefont
  {Bondani}}, \bibinfo {author} {\bibfnamefont {A.}~\bibnamefont {Allevi}},
  \bibinfo {author} {\bibfnamefont {G.}~\bibnamefont {Zambra}}, \bibinfo
  {author} {\bibfnamefont {M.~G.~A.}\ \bibnamefont {Paris}}, \ and\ \bibinfo
  {author} {\bibfnamefont {A.}~\bibnamefont {Andreoni}},\ }\href {\doibase
  10.1103/PhysRevA.76.013833} {\bibfield  {journal} {\bibinfo  {journal} {Phys.
  Rev. A}\ }\textbf {\bibinfo {volume} {76}},\ \bibinfo {pages} {013833}
  (\bibinfo {year} {2007})}\BibitemShut {NoStop}%
\bibitem [{\citenamefont {Agafonov}\ \emph {et~al.}(2010)\citenamefont
  {Agafonov}, \citenamefont {Chekhova},\ and\ \citenamefont
  {Leuchs}}]{agafonov_two-color_2010}%
  \BibitemOpen
  \bibfield  {author} {\bibinfo {author} {\bibfnamefont {I.~N.}\ \bibnamefont
  {Agafonov}}, \bibinfo {author} {\bibfnamefont {M.~V.}\ \bibnamefont
  {Chekhova}}, \ and\ \bibinfo {author} {\bibfnamefont {G.}~\bibnamefont
  {Leuchs}},\ }\href {\doibase 10.1103/PhysRevA.82.011801} {\bibfield
  {journal} {\bibinfo  {journal} {Phys. Rev. A}\ }\textbf {\bibinfo {volume}
  {82}},\ \bibinfo {pages} {011801} (\bibinfo {year} {2010})}\BibitemShut
  {NoStop}%
\bibitem [{\citenamefont {Allevi}\ and\ \citenamefont
  {Bondani}(2014)}]{allevi_statistics_2014}%
  \BibitemOpen
  \bibfield  {author} {\bibinfo {author} {\bibfnamefont {A.}~\bibnamefont
  {Allevi}}\ and\ \bibinfo {author} {\bibfnamefont {M.}~\bibnamefont
  {Bondani}},\ }\href {\doibase 10.1364/JOSAB.31.000B14} {\bibfield  {journal}
  {\bibinfo  {journal} {J. Opt. Soc. Am. B}\ }\textbf {\bibinfo {volume}
  {31}},\ \bibinfo {pages} {B14} (\bibinfo {year} {2014})}\BibitemShut
  {NoStop}%
\bibitem [{\citenamefont {Sharapova}\ \emph {et~al.}(2015)\citenamefont
  {Sharapova}, \citenamefont {Pérez}, \citenamefont {Tikhonova},\ and\
  \citenamefont {Chekhova}}]{sharapova_schmidt_2015}%
  \BibitemOpen
  \bibfield  {author} {\bibinfo {author} {\bibfnamefont {P.}~\bibnamefont
  {Sharapova}}, \bibinfo {author} {\bibfnamefont {A.~M.}\ \bibnamefont
  {Pérez}}, \bibinfo {author} {\bibfnamefont {O.~V.}\ \bibnamefont
  {Tikhonova}}, \ and\ \bibinfo {author} {\bibfnamefont {M.~V.}\ \bibnamefont
  {Chekhova}},\ }\href {\doibase 10.1103/PhysRevA.91.043816} {\bibfield
  {journal} {\bibinfo  {journal} {Phys. Rev. A}\ }\textbf {\bibinfo {volume}
  {91}},\ \bibinfo {pages} {043816} (\bibinfo {year} {2015})}\BibitemShut
  {NoStop}%
\bibitem [{\citenamefont {Eisert}\ \emph {et~al.}(2002)\citenamefont {Eisert},
  \citenamefont {Scheel},\ and\ \citenamefont
  {Plenio}}]{eisert_distilling_2002}%
  \BibitemOpen
  \bibfield  {author} {\bibinfo {author} {\bibfnamefont {J.}~\bibnamefont
  {Eisert}}, \bibinfo {author} {\bibfnamefont {S.}~\bibnamefont {Scheel}}, \
  and\ \bibinfo {author} {\bibfnamefont {M.~B.}\ \bibnamefont {Plenio}},\
  }\href {\doibase 10.1103/PhysRevLett.89.137903} {\bibfield  {journal}
  {\bibinfo  {journal} {Phys. Rev. Lett.}\ }\textbf {\bibinfo {volume} {89}},\
  \bibinfo {pages} {137903} (\bibinfo {year} {2002})}\BibitemShut {NoStop}%
\bibitem [{\citenamefont {Takahashi}\ \emph {et~al.}(2010)\citenamefont
  {Takahashi}, \citenamefont {Neergaard-Nielsen}, \citenamefont {Takeuchi},
  \citenamefont {Takeoka}, \citenamefont {Hayasaka}, \citenamefont {Furusawa},\
  and\ \citenamefont {Sasaki}}]{takahashi_entanglement_2010}%
  \BibitemOpen
  \bibfield  {author} {\bibinfo {author} {\bibfnamefont {H.}~\bibnamefont
  {Takahashi}}, \bibinfo {author} {\bibfnamefont {J.~S.}\ \bibnamefont
  {Neergaard-Nielsen}}, \bibinfo {author} {\bibfnamefont {M.}~\bibnamefont
  {Takeuchi}}, \bibinfo {author} {\bibfnamefont {M.}~\bibnamefont {Takeoka}},
  \bibinfo {author} {\bibfnamefont {K.}~\bibnamefont {Hayasaka}}, \bibinfo
  {author} {\bibfnamefont {A.}~\bibnamefont {Furusawa}}, \ and\ \bibinfo
  {author} {\bibfnamefont {M.}~\bibnamefont {Sasaki}},\ }\href {\doibase
  10.1038/nphoton.2010.1} {\bibfield  {journal} {\bibinfo  {journal} {Nat
  Photon}\ }\textbf {\bibinfo {volume} {4}},\ \bibinfo {pages} {178} (\bibinfo
  {year} {2010})}\BibitemShut {NoStop}%
\bibitem [{\citenamefont {Kurochkin}\ \emph {et~al.}(2014)\citenamefont
  {Kurochkin}, \citenamefont {Prasad},\ and\ \citenamefont
  {Lvovsky}}]{kurochkin_distillation_2014}%
  \BibitemOpen
  \bibfield  {author} {\bibinfo {author} {\bibfnamefont {Y.}~\bibnamefont
  {Kurochkin}}, \bibinfo {author} {\bibfnamefont {A.~S.}\ \bibnamefont
  {Prasad}}, \ and\ \bibinfo {author} {\bibfnamefont {A.~I.}\ \bibnamefont
  {Lvovsky}},\ }\href {\doibase 10.1103/PhysRevLett.112.070402} {\bibfield
  {journal} {\bibinfo  {journal} {Phys. Rev. Lett.}\ }\textbf {\bibinfo
  {volume} {112}},\ \bibinfo {pages} {070402} (\bibinfo {year}
  {2014})}\BibitemShut {NoStop}%
\bibitem [{\citenamefont {Ourjoumtsev}\ \emph {et~al.}(2007)\citenamefont
  {Ourjoumtsev}, \citenamefont {Jeong}, \citenamefont {Tualle-Brouri},\ and\
  \citenamefont {Grangier}}]{ourjoumtsev_generation_2007}%
  \BibitemOpen
  \bibfield  {author} {\bibinfo {author} {\bibfnamefont {A.}~\bibnamefont
  {Ourjoumtsev}}, \bibinfo {author} {\bibfnamefont {H.}~\bibnamefont {Jeong}},
  \bibinfo {author} {\bibfnamefont {R.}~\bibnamefont {Tualle-Brouri}}, \ and\
  \bibinfo {author} {\bibfnamefont {P.}~\bibnamefont {Grangier}},\ }\href
  {\doibase 10.1038/nature06054} {\bibfield  {journal} {\bibinfo  {journal}
  {Nature}\ }\textbf {\bibinfo {volume} {448}},\ \bibinfo {pages} {784}
  (\bibinfo {year} {2007})}\BibitemShut {NoStop}%
\bibitem [{\citenamefont {Gerrits}\ \emph {et~al.}(2010)\citenamefont
  {Gerrits}, \citenamefont {Glancy}, \citenamefont {Clement}, \citenamefont
  {Calkins}, \citenamefont {Lita}, \citenamefont {Miller}, \citenamefont
  {Migdall}, \citenamefont {Nam}, \citenamefont {Mirin},\ and\ \citenamefont
  {Knill}}]{gerrits_generation_2010}%
  \BibitemOpen
  \bibfield  {author} {\bibinfo {author} {\bibfnamefont {T.}~\bibnamefont
  {Gerrits}}, \bibinfo {author} {\bibfnamefont {S.}~\bibnamefont {Glancy}},
  \bibinfo {author} {\bibfnamefont {T.~S.}\ \bibnamefont {Clement}}, \bibinfo
  {author} {\bibfnamefont {B.}~\bibnamefont {Calkins}}, \bibinfo {author}
  {\bibfnamefont {A.~E.}\ \bibnamefont {Lita}}, \bibinfo {author}
  {\bibfnamefont {A.~J.}\ \bibnamefont {Miller}}, \bibinfo {author}
  {\bibfnamefont {A.~L.}\ \bibnamefont {Migdall}}, \bibinfo {author}
  {\bibfnamefont {S.~W.}\ \bibnamefont {Nam}}, \bibinfo {author} {\bibfnamefont
  {R.~P.}\ \bibnamefont {Mirin}}, \ and\ \bibinfo {author} {\bibfnamefont
  {E.}~\bibnamefont {Knill}},\ }\href {\doibase 10.1103/PhysRevA.82.031802}
  {\bibfield  {journal} {\bibinfo  {journal} {Phys. Rev. A}\ }\textbf {\bibinfo
  {volume} {82}},\ \bibinfo {pages} {031802} (\bibinfo {year}
  {2010})}\BibitemShut {NoStop}%
\bibitem [{\citenamefont {Lita}\ \emph {et~al.}(2008)\citenamefont {Lita},
  \citenamefont {Miller},\ and\ \citenamefont {Nam}}]{lita_counting_2008}%
  \BibitemOpen
  \bibfield  {author} {\bibinfo {author} {\bibfnamefont {A.~E.}\ \bibnamefont
  {Lita}}, \bibinfo {author} {\bibfnamefont {A.~J.}\ \bibnamefont {Miller}}, \
  and\ \bibinfo {author} {\bibfnamefont {S.~W.}\ \bibnamefont {Nam}},\ }\href
  {\doibase 10.1364/OE.16.003032} {\bibfield  {journal} {\bibinfo  {journal}
  {Opt. Express}\ }\textbf {\bibinfo {volume} {16}},\ \bibinfo {pages} {3032}
  (\bibinfo {year} {2008})}\BibitemShut {NoStop}%
\bibitem [{\citenamefont {Marsili}\ \emph {et~al.}(2013)\citenamefont
  {Marsili}, \citenamefont {Verma}, \citenamefont {Stern}, \citenamefont
  {Harrington}, \citenamefont {Lita}, \citenamefont {Gerrits}, \citenamefont
  {Vayshenker}, \citenamefont {Baek}, \citenamefont {Shaw}, \citenamefont
  {Mirin},\ and\ \citenamefont {Nam}}]{marsili_detecting_2013}%
  \BibitemOpen
  \bibfield  {author} {\bibinfo {author} {\bibfnamefont {F.}~\bibnamefont
  {Marsili}}, \bibinfo {author} {\bibfnamefont {V.~B.}\ \bibnamefont {Verma}},
  \bibinfo {author} {\bibfnamefont {J.~A.}\ \bibnamefont {Stern}}, \bibinfo
  {author} {\bibfnamefont {S.}~\bibnamefont {Harrington}}, \bibinfo {author}
  {\bibfnamefont {A.~E.}\ \bibnamefont {Lita}}, \bibinfo {author}
  {\bibfnamefont {T.}~\bibnamefont {Gerrits}}, \bibinfo {author} {\bibfnamefont
  {I.}~\bibnamefont {Vayshenker}}, \bibinfo {author} {\bibfnamefont
  {B.}~\bibnamefont {Baek}}, \bibinfo {author} {\bibfnamefont {M.~D.}\
  \bibnamefont {Shaw}}, \bibinfo {author} {\bibfnamefont {R.~P.}\ \bibnamefont
  {Mirin}}, \ and\ \bibinfo {author} {\bibfnamefont {S.~W.}\ \bibnamefont
  {Nam}},\ }\href {\doibase 10.1038/nphoton.2013.13} {\bibfield  {journal}
  {\bibinfo  {journal} {Nat Photon}\ }\textbf {\bibinfo {volume} {7}},\
  \bibinfo {pages} {210} (\bibinfo {year} {2013})}\BibitemShut {NoStop}%
\bibitem [{\citenamefont {Pérez}\ \emph {et~al.}(2014)\citenamefont {Pérez},
  \citenamefont {Iskhakov}, \citenamefont {Sharapova}, \citenamefont {Lemieux},
  \citenamefont {Tikhonova}, \citenamefont {Chekhova},\ and\ \citenamefont
  {Leuchs}}]{perez_bright_2014}%
  \BibitemOpen
  \bibfield  {author} {\bibinfo {author} {\bibfnamefont {A.~M.}\ \bibnamefont
  {Pérez}}, \bibinfo {author} {\bibfnamefont {T.~S.}\ \bibnamefont
  {Iskhakov}}, \bibinfo {author} {\bibfnamefont {P.}~\bibnamefont {Sharapova}},
  \bibinfo {author} {\bibfnamefont {S.}~\bibnamefont {Lemieux}}, \bibinfo
  {author} {\bibfnamefont {O.~V.}\ \bibnamefont {Tikhonova}}, \bibinfo {author}
  {\bibfnamefont {M.~V.}\ \bibnamefont {Chekhova}}, \ and\ \bibinfo {author}
  {\bibfnamefont {G.}~\bibnamefont {Leuchs}},\ }\href {\doibase
  10.1364/OL.39.002403} {\bibfield  {journal} {\bibinfo  {journal} {Opt.
  Lett.}\ }\textbf {\bibinfo {volume} {39}},\ \bibinfo {pages} {2403} (\bibinfo
  {year} {2014})}\BibitemShut {NoStop}%
\bibitem [{\citenamefont {Brecht}\ \emph {et~al.}(2014)\citenamefont {Brecht},
  \citenamefont {Eckstein}, \citenamefont {Ricken}, \citenamefont {Quiring},
  \citenamefont {Suche}, \citenamefont {Sansoni},\ and\ \citenamefont
  {Silberhorn}}]{brecht_demonstration_2014}%
  \BibitemOpen
  \bibfield  {author} {\bibinfo {author} {\bibfnamefont {B.}~\bibnamefont
  {Brecht}}, \bibinfo {author} {\bibfnamefont {A.}~\bibnamefont {Eckstein}},
  \bibinfo {author} {\bibfnamefont {R.}~\bibnamefont {Ricken}}, \bibinfo
  {author} {\bibfnamefont {V.}~\bibnamefont {Quiring}}, \bibinfo {author}
  {\bibfnamefont {H.}~\bibnamefont {Suche}}, \bibinfo {author} {\bibfnamefont
  {L.}~\bibnamefont {Sansoni}}, \ and\ \bibinfo {author} {\bibfnamefont
  {C.}~\bibnamefont {Silberhorn}},\ }\href {\doibase
  10.1103/PhysRevA.90.030302} {\bibfield  {journal} {\bibinfo  {journal} {Phys.
  Rev. A}\ }\textbf {\bibinfo {volume} {90}},\ \bibinfo {pages} {030302}
  (\bibinfo {year} {2014})}\BibitemShut {NoStop}%
\bibitem [{\citenamefont {Christ}\ \emph {et~al.}(2014)\citenamefont {Christ},
  \citenamefont {Lupo}, \citenamefont {Reichelt}, \citenamefont {Meier},\ and\
  \citenamefont {Silberhorn}}]{christ_theory_2014}%
  \BibitemOpen
  \bibfield  {author} {\bibinfo {author} {\bibfnamefont {A.}~\bibnamefont
  {Christ}}, \bibinfo {author} {\bibfnamefont {C.}~\bibnamefont {Lupo}},
  \bibinfo {author} {\bibfnamefont {M.}~\bibnamefont {Reichelt}}, \bibinfo
  {author} {\bibfnamefont {T.}~\bibnamefont {Meier}}, \ and\ \bibinfo {author}
  {\bibfnamefont {C.}~\bibnamefont {Silberhorn}},\ }\href {\doibase
  10.1103/PhysRevA.90.023823} {\bibfield  {journal} {\bibinfo  {journal} {Phys.
  Rev. A}\ }\textbf {\bibinfo {volume} {90}},\ \bibinfo {pages} {023823}
  (\bibinfo {year} {2014})}\BibitemShut {NoStop}%
\bibitem [{\citenamefont {Grice}\ \emph {et~al.}(2001)\citenamefont {Grice},
  \citenamefont {U’Ren},\ and\ \citenamefont
  {Walmsley}}]{grice_eliminating_2001}%
  \BibitemOpen
  \bibfield  {author} {\bibinfo {author} {\bibfnamefont {W.~P.}\ \bibnamefont
  {Grice}}, \bibinfo {author} {\bibfnamefont {A.~B.}\ \bibnamefont {U’Ren}},
  \ and\ \bibinfo {author} {\bibfnamefont {I.~A.}\ \bibnamefont {Walmsley}},\
  }\href {\doibase 10.1103/PhysRevA.64.063815} {\bibfield  {journal} {\bibinfo
  {journal} {Phys. Rev. A}\ }\textbf {\bibinfo {volume} {64}},\ \bibinfo
  {pages} {063815} (\bibinfo {year} {2001})}\BibitemShut {NoStop}%
\bibitem [{\citenamefont {U'Ren}\ \emph {et~al.}(2005)\citenamefont {U'Ren},
  \citenamefont {Silberhorn}, \citenamefont {Banaszek}, \citenamefont
  {Walmsley}, \citenamefont {Erdmann}, \citenamefont {Grice},\ and\
  \citenamefont {Raymer}}]{uren_generation_2005}%
  \BibitemOpen
  \bibfield  {author} {\bibinfo {author} {\bibfnamefont {A.~B.}\ \bibnamefont
  {U'Ren}}, \bibinfo {author} {\bibfnamefont {C.}~\bibnamefont {Silberhorn}},
  \bibinfo {author} {\bibfnamefont {K.}~\bibnamefont {Banaszek}}, \bibinfo
  {author} {\bibfnamefont {I.~A.}\ \bibnamefont {Walmsley}}, \bibinfo {author}
  {\bibfnamefont {R.}~\bibnamefont {Erdmann}}, \bibinfo {author} {\bibfnamefont
  {W.~P.}\ \bibnamefont {Grice}}, \ and\ \bibinfo {author} {\bibfnamefont
  {M.~G.}\ \bibnamefont {Raymer}},\ }\href@noop {} {\bibfield  {journal}
  {\bibinfo  {journal} {Laser Phys.}\ }\textbf {\bibinfo {volume} {15}},\
  \bibinfo {pages} {146} (\bibinfo {year} {2005})}\BibitemShut {NoStop}%
\bibitem [{\citenamefont {Mosley}\ \emph {et~al.}(2008)\citenamefont {Mosley},
  \citenamefont {Lundeen}, \citenamefont {Smith}, \citenamefont {Wasylczyk},
  \citenamefont {U’Ren}, \citenamefont {Silberhorn},\ and\ \citenamefont
  {Walmsley}}]{Mosley_heralded_2008}%
  \BibitemOpen
  \bibfield  {author} {\bibinfo {author} {\bibfnamefont {P.~J.}\ \bibnamefont
  {Mosley}}, \bibinfo {author} {\bibfnamefont {J.~S.}\ \bibnamefont {Lundeen}},
  \bibinfo {author} {\bibfnamefont {B.~J.}\ \bibnamefont {Smith}}, \bibinfo
  {author} {\bibfnamefont {P.}~\bibnamefont {Wasylczyk}}, \bibinfo {author}
  {\bibfnamefont {A.~B.}\ \bibnamefont {U’Ren}}, \bibinfo {author}
  {\bibfnamefont {C.}~\bibnamefont {Silberhorn}}, \ and\ \bibinfo {author}
  {\bibfnamefont {I.~A.}\ \bibnamefont {Walmsley}},\ }\href {\doibase
  10.1103/PhysRevLett.100.133601} {\bibfield  {journal} {\bibinfo  {journal}
  {Phys. Rev. Lett.}\ }\textbf {\bibinfo {volume} {100}},\ \bibinfo {pages}
  {133601} (\bibinfo {year} {2008})}\BibitemShut {NoStop}%
\bibitem [{\citenamefont {Eckstein}\ \emph {et~al.}(2011)\citenamefont
  {Eckstein}, \citenamefont {Christ}, \citenamefont {Mosley},\ and\
  \citenamefont {Silberhorn}}]{eckstein_highly_2011}%
  \BibitemOpen
  \bibfield  {author} {\bibinfo {author} {\bibfnamefont {A.}~\bibnamefont
  {Eckstein}}, \bibinfo {author} {\bibfnamefont {A.}~\bibnamefont {Christ}},
  \bibinfo {author} {\bibfnamefont {P.~J.}\ \bibnamefont {Mosley}}, \ and\
  \bibinfo {author} {\bibfnamefont {C.}~\bibnamefont {Silberhorn}},\ }\href
  {\doibase 10.1103/PhysRevLett.106.013603} {\bibfield  {journal} {\bibinfo
  {journal} {Phys. Rev. Lett.}\ }\textbf {\bibinfo {volume} {106}},\ \bibinfo
  {pages} {013603} (\bibinfo {year} {2011})}\BibitemShut {NoStop}%
\bibitem [{\citenamefont {Quesada}\ and\ \citenamefont
  {Sipe}(2014)}]{quesada_effects_2014}%
  \BibitemOpen
  \bibfield  {author} {\bibinfo {author} {\bibfnamefont {N.}~\bibnamefont
  {Quesada}}\ and\ \bibinfo {author} {\bibfnamefont {J.~E.}\ \bibnamefont
  {Sipe}},\ }\href {\doibase 10.1103/PhysRevA.90.063840} {\bibfield  {journal}
  {\bibinfo  {journal} {Phys. Rev. A}\ }\textbf {\bibinfo {volume} {90}},\
  \bibinfo {pages} {063840} (\bibinfo {year} {2014})}\BibitemShut {NoStop}%
\bibitem [{\citenamefont {Sundheimer}\ \emph {et~al.}(1993)\citenamefont
  {Sundheimer}, \citenamefont {Bosshard}, \citenamefont {Van~Stryland},
  \citenamefont {Stegeman},\ and\ \citenamefont
  {Bierlein}}]{sundheimer_large_1993}%
  \BibitemOpen
  \bibfield  {author} {\bibinfo {author} {\bibfnamefont {M.~L.}\ \bibnamefont
  {Sundheimer}}, \bibinfo {author} {\bibfnamefont {C.}~\bibnamefont
  {Bosshard}}, \bibinfo {author} {\bibfnamefont {E.~W.}\ \bibnamefont
  {Van~Stryland}}, \bibinfo {author} {\bibfnamefont {G.~I.}\ \bibnamefont
  {Stegeman}}, \ and\ \bibinfo {author} {\bibfnamefont {J.~D.}\ \bibnamefont
  {Bierlein}},\ }\href {\doibase 10.1364/OL.18.001397} {\bibfield  {journal}
  {\bibinfo  {journal} {Optics Letters}\ }\textbf {\bibinfo {volume} {18}},\
  \bibinfo {pages} {1397} (\bibinfo {year} {1993})}\BibitemShut {NoStop}%
\bibitem [{\citenamefont {Tyminski}(1991)}]{tyminski_photorefractive_1991}%
  \BibitemOpen
  \bibfield  {author} {\bibinfo {author} {\bibfnamefont {J.~K.}\ \bibnamefont
  {Tyminski}},\ }\href {\doibase 10.1063/1.350194} {\bibfield  {journal}
  {\bibinfo  {journal} {Journal of Applied Physics}\ }\textbf {\bibinfo
  {volume} {70}},\ \bibinfo {pages} {5570} (\bibinfo {year}
  {1991})}\BibitemShut {NoStop}%
\bibitem [{\citenamefont {Wang}\ \emph {et~al.}(2000)\citenamefont {Wang},
  \citenamefont {Wei}, \citenamefont {Liu}, \citenamefont {Yin}, \citenamefont
  {Hu}, \citenamefont {Shao},\ and\ \citenamefont {Jiang}}]{wang_survey_2000}%
  \BibitemOpen
  \bibfield  {author} {\bibinfo {author} {\bibfnamefont {J.}~\bibnamefont
  {Wang}}, \bibinfo {author} {\bibfnamefont {J.}~\bibnamefont {Wei}}, \bibinfo
  {author} {\bibfnamefont {Y.}~\bibnamefont {Liu}}, \bibinfo {author}
  {\bibfnamefont {X.}~\bibnamefont {Yin}}, \bibinfo {author} {\bibfnamefont
  {X.}~\bibnamefont {Hu}}, \bibinfo {author} {\bibfnamefont {Z.}~\bibnamefont
  {Shao}}, \ and\ \bibinfo {author} {\bibfnamefont {M.}~\bibnamefont {Jiang}},\
  }\href {\doibase 10.1016/S0960-8974(00)00007-3} {\bibfield  {journal}
  {\bibinfo  {journal} {Progress in Crystal Growth and Characterization of
  Materials}\ }\bibinfo {series} {Symposium {L} {Fifth} {IUMRS} {International}
  {Conference} on {Advanced} {Materials} {IUMRS}-{ICAM}\_99},\ \textbf
  {\bibinfo {volume} {40}},\ \bibinfo {pages} {3} (\bibinfo {year}
  {2000})}\BibitemShut {NoStop}%
\bibitem [{\citenamefont {Fiorentino}\ \emph {et~al.}(2007)\citenamefont
  {Fiorentino}, \citenamefont {Spillane}, \citenamefont {Beausoleil},
  \citenamefont {Roberts}, \citenamefont {Battle},\ and\ \citenamefont
  {Munro}}]{fiorentino_spontaneous_2007}%
  \BibitemOpen
  \bibfield  {author} {\bibinfo {author} {\bibfnamefont {M.}~\bibnamefont
  {Fiorentino}}, \bibinfo {author} {\bibfnamefont {S.~M.}\ \bibnamefont
  {Spillane}}, \bibinfo {author} {\bibfnamefont {R.~G.}\ \bibnamefont
  {Beausoleil}}, \bibinfo {author} {\bibfnamefont {T.~D.}\ \bibnamefont
  {Roberts}}, \bibinfo {author} {\bibfnamefont {P.}~\bibnamefont {Battle}}, \
  and\ \bibinfo {author} {\bibfnamefont {M.~W.}\ \bibnamefont {Munro}},\ }\href
  {\doibase 10.1364/OE.15.007479} {\bibfield  {journal} {\bibinfo  {journal}
  {Opt. Express}\ }\textbf {\bibinfo {volume} {15}},\ \bibinfo {pages} {7479}
  (\bibinfo {year} {2007})}\BibitemShut {NoStop}%
\bibitem [{\citenamefont {Harder}\ \emph {et~al.}(2013)\citenamefont {Harder},
  \citenamefont {Ansari}, \citenamefont {Brecht}, \citenamefont {Dirmeier},
  \citenamefont {Marquardt},\ and\ \citenamefont
  {Silberhorn}}]{harder_optimized_2013}%
  \BibitemOpen
  \bibfield  {author} {\bibinfo {author} {\bibfnamefont {G.}~\bibnamefont
  {Harder}}, \bibinfo {author} {\bibfnamefont {V.}~\bibnamefont {Ansari}},
  \bibinfo {author} {\bibfnamefont {B.}~\bibnamefont {Brecht}}, \bibinfo
  {author} {\bibfnamefont {T.}~\bibnamefont {Dirmeier}}, \bibinfo {author}
  {\bibfnamefont {C.}~\bibnamefont {Marquardt}}, \ and\ \bibinfo {author}
  {\bibfnamefont {C.}~\bibnamefont {Silberhorn}},\ }\href {\doibase
  10.1364/OE.21.013975} {\bibfield  {journal} {\bibinfo  {journal} {Opt.
  Express}\ }\textbf {\bibinfo {volume} {21}},\ \bibinfo {pages} {13975}
  (\bibinfo {year} {2013})}\BibitemShut {NoStop}%
\bibitem [{\citenamefont {Gerrits}\ \emph {et~al.}(2012)\citenamefont
  {Gerrits}, \citenamefont {Calkins}, \citenamefont {Tomlin}, \citenamefont
  {Lita}, \citenamefont {Migdall}, \citenamefont {Mirin},\ and\ \citenamefont
  {Nam}}]{gerrits_extending_2012}%
  \BibitemOpen
  \bibfield  {author} {\bibinfo {author} {\bibfnamefont {T.}~\bibnamefont
  {Gerrits}}, \bibinfo {author} {\bibfnamefont {B.}~\bibnamefont {Calkins}},
  \bibinfo {author} {\bibfnamefont {N.}~\bibnamefont {Tomlin}}, \bibinfo
  {author} {\bibfnamefont {A.~E.}\ \bibnamefont {Lita}}, \bibinfo {author}
  {\bibfnamefont {A.}~\bibnamefont {Migdall}}, \bibinfo {author} {\bibfnamefont
  {R.}~\bibnamefont {Mirin}}, \ and\ \bibinfo {author} {\bibfnamefont {S.~W.}\
  \bibnamefont {Nam}},\ }\href {\doibase 10.1364/OE.20.023798} {\bibfield
  {journal} {\bibinfo  {journal} {Opt. Express}\ }\textbf {\bibinfo {volume}
  {20}},\ \bibinfo {pages} {23798} (\bibinfo {year} {2012})}\BibitemShut
  {NoStop}%
\bibitem [{\citenamefont {Loudon}(2000)}]{loudon_quantum_2000}%
  \BibitemOpen
  \bibfield  {author} {\bibinfo {author} {\bibfnamefont {R.}~\bibnamefont
  {Loudon}},\ }\href@noop {} {\emph {\bibinfo {title} {The {Quantum} {Theory}
  of {Light}}}}\ (\bibinfo  {publisher} {OUP Oxford},\ \bibinfo {year}
  {2000})\BibitemShut {NoStop}%
\bibitem [{\citenamefont {Eberly}(2006)}]{eberly_schmidt_2006}%
  \BibitemOpen
  \bibfield  {author} {\bibinfo {author} {\bibfnamefont {J.~H.}\ \bibnamefont
  {Eberly}},\ }\href {\doibase 10.1134/S1054660X06060041} {\bibfield  {journal}
  {\bibinfo  {journal} {Laser Phys.}\ }\textbf {\bibinfo {volume} {16}},\
  \bibinfo {pages} {921} (\bibinfo {year} {2006})}\BibitemShut {NoStop}%
\bibitem [{\citenamefont {Christ}\ \emph {et~al.}(2011)\citenamefont {Christ},
  \citenamefont {Laiho}, \citenamefont {Eckstein}, \citenamefont {Cassemiro},\
  and\ \citenamefont {Silberhorn}}]{christ_probing_2011}%
  \BibitemOpen
  \bibfield  {author} {\bibinfo {author} {\bibfnamefont {A.}~\bibnamefont
  {Christ}}, \bibinfo {author} {\bibfnamefont {K.}~\bibnamefont {Laiho}},
  \bibinfo {author} {\bibfnamefont {A.}~\bibnamefont {Eckstein}}, \bibinfo
  {author} {\bibfnamefont {K.~N.}\ \bibnamefont {Cassemiro}}, \ and\ \bibinfo
  {author} {\bibfnamefont {C.}~\bibnamefont {Silberhorn}},\ }\href {\doibase
  10.1088/1367-2630/13/3/033027} {\bibfield  {journal} {\bibinfo  {journal}
  {New J. Phys.}\ }\textbf {\bibinfo {volume} {13}},\ \bibinfo {pages} {033027}
  (\bibinfo {year} {2011})}\BibitemShut {NoStop}%
\bibitem [{\citenamefont {Klyshko}(1980)}]{klyshko_use_1980}%
  \BibitemOpen
  \bibfield  {author} {\bibinfo {author} {\bibfnamefont {D.~N.}\ \bibnamefont
  {Klyshko}},\ }\href {\doibase 10.1070/QE1980v010n09ABEH010660} {\bibfield
  {journal} {\bibinfo  {journal} {Sov. J. Quantum Electron.}\ }\textbf
  {\bibinfo {volume} {10}},\ \bibinfo {pages} {1112} (\bibinfo {year}
  {1980})}\BibitemShut {NoStop}%
\bibitem [{\citenamefont {Worsley}\ \emph {et~al.}(2009)\citenamefont
  {Worsley}, \citenamefont {Coldenstrodt-Ronge}, \citenamefont {Lundeen},
  \citenamefont {Mosley}, \citenamefont {Smith}, \citenamefont {Puentes},
  \citenamefont {Thomas-Peter},\ and\ \citenamefont
  {Walmsley}}]{worsley_absolute_2009}%
  \BibitemOpen
  \bibfield  {author} {\bibinfo {author} {\bibfnamefont {A.~P.}\ \bibnamefont
  {Worsley}}, \bibinfo {author} {\bibfnamefont {H.~B.}\ \bibnamefont
  {Coldenstrodt-Ronge}}, \bibinfo {author} {\bibfnamefont {J.~S.}\ \bibnamefont
  {Lundeen}}, \bibinfo {author} {\bibfnamefont {P.~J.}\ \bibnamefont {Mosley}},
  \bibinfo {author} {\bibfnamefont {B.~J.}\ \bibnamefont {Smith}}, \bibinfo
  {author} {\bibfnamefont {G.}~\bibnamefont {Puentes}}, \bibinfo {author}
  {\bibfnamefont {N.}~\bibnamefont {Thomas-Peter}}, \ and\ \bibinfo {author}
  {\bibfnamefont {I.~A.}\ \bibnamefont {Walmsley}},\ }\href {\doibase
  10.1364/OE.17.004397} {\bibfield  {journal} {\bibinfo  {journal} {Opt.
  Express}\ }\textbf {\bibinfo {volume} {17}},\ \bibinfo {pages} {4397}
  (\bibinfo {year} {2009})}\BibitemShut {NoStop}%
\bibitem [{\citenamefont {Giustina}\ \emph {et~al.}(2013)\citenamefont
  {Giustina}, \citenamefont {Mech}, \citenamefont {Ramelow}, \citenamefont
  {Wittmann}, \citenamefont {Kofler}, \citenamefont {Beyer}, \citenamefont
  {Lita}, \citenamefont {Calkins}, \citenamefont {Gerrits}, \citenamefont
  {Nam}, \citenamefont {Ursin},\ and\ \citenamefont
  {Zeilinger}}]{giustina_bell_2013}%
  \BibitemOpen
  \bibfield  {author} {\bibinfo {author} {\bibfnamefont {M.}~\bibnamefont
  {Giustina}}, \bibinfo {author} {\bibfnamefont {A.}~\bibnamefont {Mech}},
  \bibinfo {author} {\bibfnamefont {S.}~\bibnamefont {Ramelow}}, \bibinfo
  {author} {\bibfnamefont {B.}~\bibnamefont {Wittmann}}, \bibinfo {author}
  {\bibfnamefont {J.}~\bibnamefont {Kofler}}, \bibinfo {author} {\bibfnamefont
  {J.}~\bibnamefont {Beyer}}, \bibinfo {author} {\bibfnamefont
  {A.}~\bibnamefont {Lita}}, \bibinfo {author} {\bibfnamefont {B.}~\bibnamefont
  {Calkins}}, \bibinfo {author} {\bibfnamefont {T.}~\bibnamefont {Gerrits}},
  \bibinfo {author} {\bibfnamefont {S.~W.}\ \bibnamefont {Nam}}, \bibinfo
  {author} {\bibfnamefont {R.}~\bibnamefont {Ursin}}, \ and\ \bibinfo {author}
  {\bibfnamefont {A.}~\bibnamefont {Zeilinger}},\ }\href {\doibase
  10.1038/nature12012} {\bibfield  {journal} {\bibinfo  {journal} {Nature}\
  }\textbf {\bibinfo {volume} {497}},\ \bibinfo {pages} {227} (\bibinfo {year}
  {2013})}\BibitemShut {NoStop}%
\bibitem [{\citenamefont {Christensen}\ \emph {et~al.}(2013)\citenamefont
  {Christensen}, \citenamefont {McCusker}, \citenamefont {Altepeter},
  \citenamefont {Calkins}, \citenamefont {Gerrits}, \citenamefont {Lita},
  \citenamefont {Miller}, \citenamefont {Shalm}, \citenamefont {Zhang},
  \citenamefont {Nam}, \citenamefont {Brunner}, \citenamefont {Lim},
  \citenamefont {Gisin},\ and\ \citenamefont
  {Kwiat}}]{christensen_detection-loophole-free_2013}%
  \BibitemOpen
  \bibfield  {author} {\bibinfo {author} {\bibfnamefont {B.~G.}\ \bibnamefont
  {Christensen}}, \bibinfo {author} {\bibfnamefont {K.~T.}\ \bibnamefont
  {McCusker}}, \bibinfo {author} {\bibfnamefont {J.~B.}\ \bibnamefont
  {Altepeter}}, \bibinfo {author} {\bibfnamefont {B.}~\bibnamefont {Calkins}},
  \bibinfo {author} {\bibfnamefont {T.}~\bibnamefont {Gerrits}}, \bibinfo
  {author} {\bibfnamefont {A.~E.}\ \bibnamefont {Lita}}, \bibinfo {author}
  {\bibfnamefont {A.}~\bibnamefont {Miller}}, \bibinfo {author} {\bibfnamefont
  {L.~K.}\ \bibnamefont {Shalm}}, \bibinfo {author} {\bibfnamefont
  {Y.}~\bibnamefont {Zhang}}, \bibinfo {author} {\bibfnamefont {S.~W.}\
  \bibnamefont {Nam}}, \bibinfo {author} {\bibfnamefont {N.}~\bibnamefont
  {Brunner}}, \bibinfo {author} {\bibfnamefont {C.~C.~W.}\ \bibnamefont {Lim}},
  \bibinfo {author} {\bibfnamefont {N.}~\bibnamefont {Gisin}}, \ and\ \bibinfo
  {author} {\bibfnamefont {P.~G.}\ \bibnamefont {Kwiat}},\ }\href {\doibase
  10.1103/PhysRevLett.111.130406} {\bibfield  {journal} {\bibinfo  {journal}
  {Phys. Rev. Lett.}\ }\textbf {\bibinfo {volume} {111}},\ \bibinfo {pages}
  {130406} (\bibinfo {year} {2013})}\BibitemShut {NoStop}%
\bibitem [{\citenamefont {Ramelow}\ \emph {et~al.}(2013)\citenamefont
  {Ramelow}, \citenamefont {Mech}, \citenamefont {Giustina}, \citenamefont
  {Gröblacher}, \citenamefont {Wieczorek}, \citenamefont {Beyer},
  \citenamefont {Lita}, \citenamefont {Calkins}, \citenamefont {Gerrits},
  \citenamefont {Nam}, \citenamefont {Zeilinger},\ and\ \citenamefont
  {Ursin}}]{ramelow_highly_2013}%
  \BibitemOpen
  \bibfield  {author} {\bibinfo {author} {\bibfnamefont {S.}~\bibnamefont
  {Ramelow}}, \bibinfo {author} {\bibfnamefont {A.}~\bibnamefont {Mech}},
  \bibinfo {author} {\bibfnamefont {M.}~\bibnamefont {Giustina}}, \bibinfo
  {author} {\bibfnamefont {S.}~\bibnamefont {Gröblacher}}, \bibinfo {author}
  {\bibfnamefont {W.}~\bibnamefont {Wieczorek}}, \bibinfo {author}
  {\bibfnamefont {J.}~\bibnamefont {Beyer}}, \bibinfo {author} {\bibfnamefont
  {A.}~\bibnamefont {Lita}}, \bibinfo {author} {\bibfnamefont {B.}~\bibnamefont
  {Calkins}}, \bibinfo {author} {\bibfnamefont {T.}~\bibnamefont {Gerrits}},
  \bibinfo {author} {\bibfnamefont {S.~W.}\ \bibnamefont {Nam}}, \bibinfo
  {author} {\bibfnamefont {A.}~\bibnamefont {Zeilinger}}, \ and\ \bibinfo
  {author} {\bibfnamefont {R.}~\bibnamefont {Ursin}},\ }\href {\doibase
  10.1364/OE.21.006707} {\bibfield  {journal} {\bibinfo  {journal} {Optics
  Express}\ }\textbf {\bibinfo {volume} {21}},\ \bibinfo {pages} {6707}
  (\bibinfo {year} {2013})}\BibitemShut {NoStop}%
\bibitem [{\citenamefont {Dixon}\ \emph {et~al.}(2014)\citenamefont {Dixon},
  \citenamefont {Rosenberg}, \citenamefont {Stelmakh}, \citenamefont {Grein},
  \citenamefont {Bennink}, \citenamefont {Dauler}, \citenamefont {Kerman},
  \citenamefont {Molnar},\ and\ \citenamefont {Wong}}]{dixon_heralding_2014}%
  \BibitemOpen
  \bibfield  {author} {\bibinfo {author} {\bibfnamefont {P.~B.}\ \bibnamefont
  {Dixon}}, \bibinfo {author} {\bibfnamefont {D.}~\bibnamefont {Rosenberg}},
  \bibinfo {author} {\bibfnamefont {V.}~\bibnamefont {Stelmakh}}, \bibinfo
  {author} {\bibfnamefont {M.~E.}\ \bibnamefont {Grein}}, \bibinfo {author}
  {\bibfnamefont {R.~S.}\ \bibnamefont {Bennink}}, \bibinfo {author}
  {\bibfnamefont {E.~A.}\ \bibnamefont {Dauler}}, \bibinfo {author}
  {\bibfnamefont {A.~J.}\ \bibnamefont {Kerman}}, \bibinfo {author}
  {\bibfnamefont {R.~J.}\ \bibnamefont {Molnar}}, \ and\ \bibinfo {author}
  {\bibfnamefont {F.~N.~C.}\ \bibnamefont {Wong}},\ }\href {\doibase
  10.1103/PhysRevA.90.043804} {\bibfield  {journal} {\bibinfo  {journal} {Phys.
  Rev. A}\ }\textbf {\bibinfo {volume} {90}},\ \bibinfo {pages} {043804}
  (\bibinfo {year} {2014})}\BibitemShut {NoStop}%
\bibitem [{\citenamefont {Eto}\ \emph {et~al.}(2011)\citenamefont {Eto},
  \citenamefont {Koshio}, \citenamefont {Ohshiro}, \citenamefont {Sakurai},
  \citenamefont {Horie}, \citenamefont {Hirano},\ and\ \citenamefont
  {Sasaki}}]{eto_efficient_2011}%
  \BibitemOpen
  \bibfield  {author} {\bibinfo {author} {\bibfnamefont {Y.}~\bibnamefont
  {Eto}}, \bibinfo {author} {\bibfnamefont {A.}~\bibnamefont {Koshio}},
  \bibinfo {author} {\bibfnamefont {A.}~\bibnamefont {Ohshiro}}, \bibinfo
  {author} {\bibfnamefont {J.}~\bibnamefont {Sakurai}}, \bibinfo {author}
  {\bibfnamefont {K.}~\bibnamefont {Horie}}, \bibinfo {author} {\bibfnamefont
  {T.}~\bibnamefont {Hirano}}, \ and\ \bibinfo {author} {\bibfnamefont
  {M.}~\bibnamefont {Sasaki}},\ }\href {\doibase 10.1364/OL.36.004653}
  {\bibfield  {journal} {\bibinfo  {journal} {Optics Letters}\ }\textbf
  {\bibinfo {volume} {36}},\ \bibinfo {pages} {4653} (\bibinfo {year}
  {2011})}\BibitemShut {NoStop}%
\bibitem [{\citenamefont {Eberle}\ \emph {et~al.}(2010)\citenamefont {Eberle},
  \citenamefont {Steinlechner}, \citenamefont {Bauchrowitz}, \citenamefont
  {Händchen}, \citenamefont {Vahlbruch}, \citenamefont {Mehmet}, \citenamefont
  {Müller-Ebhardt},\ and\ \citenamefont {Schnabel}}]{eberle_quantum_2010}%
  \BibitemOpen
  \bibfield  {author} {\bibinfo {author} {\bibfnamefont {T.}~\bibnamefont
  {Eberle}}, \bibinfo {author} {\bibfnamefont {S.}~\bibnamefont
  {Steinlechner}}, \bibinfo {author} {\bibfnamefont {J.}~\bibnamefont
  {Bauchrowitz}}, \bibinfo {author} {\bibfnamefont {V.}~\bibnamefont
  {Händchen}}, \bibinfo {author} {\bibfnamefont {H.}~\bibnamefont
  {Vahlbruch}}, \bibinfo {author} {\bibfnamefont {M.}~\bibnamefont {Mehmet}},
  \bibinfo {author} {\bibfnamefont {H.}~\bibnamefont {Müller-Ebhardt}}, \ and\
  \bibinfo {author} {\bibfnamefont {R.}~\bibnamefont {Schnabel}},\ }\href
  {\doibase 10.1103/PhysRevLett.104.251102} {\bibfield  {journal} {\bibinfo
  {journal} {Phys. Rev. Lett.}\ }\textbf {\bibinfo {volume} {104}},\ \bibinfo
  {pages} {251102} (\bibinfo {year} {2010})}\BibitemShut {NoStop}%
\bibitem [{\citenamefont {Alferness}\ and\ \citenamefont
  {Buhl}(1984)}]{alferness_low-cross-talk_1984}%
  \BibitemOpen
  \bibfield  {author} {\bibinfo {author} {\bibfnamefont {R.~C.}\ \bibnamefont
  {Alferness}}\ and\ \bibinfo {author} {\bibfnamefont {L.~L.}\ \bibnamefont
  {Buhl}},\ }\href {\doibase 10.1364/OL.9.000140} {\bibfield  {journal}
  {\bibinfo  {journal} {Opt. Lett.}\ }\textbf {\bibinfo {volume} {9}},\
  \bibinfo {pages} {140} (\bibinfo {year} {1984})}\BibitemShut {NoStop}%
\bibitem [{\citenamefont {Calkins}\ \emph {et~al.}(2013)\citenamefont
  {Calkins}, \citenamefont {Mennea}, \citenamefont {Lita}, \citenamefont
  {Metcalf}, \citenamefont {Kolthammer}, \citenamefont {Lamas-Linares},
  \citenamefont {Spring}, \citenamefont {Humphreys}, \citenamefont {Mirin},
  \citenamefont {Gates}, \citenamefont {Smith}, \citenamefont {Walmsley},
  \citenamefont {Gerrits},\ and\ \citenamefont {Nam}}]{calkins_high_2013}%
  \BibitemOpen
  \bibfield  {author} {\bibinfo {author} {\bibfnamefont {B.}~\bibnamefont
  {Calkins}}, \bibinfo {author} {\bibfnamefont {P.~L.}\ \bibnamefont {Mennea}},
  \bibinfo {author} {\bibfnamefont {A.~E.}\ \bibnamefont {Lita}}, \bibinfo
  {author} {\bibfnamefont {B.~J.}\ \bibnamefont {Metcalf}}, \bibinfo {author}
  {\bibfnamefont {W.~S.}\ \bibnamefont {Kolthammer}}, \bibinfo {author}
  {\bibfnamefont {A.}~\bibnamefont {Lamas-Linares}}, \bibinfo {author}
  {\bibfnamefont {J.~B.}\ \bibnamefont {Spring}}, \bibinfo {author}
  {\bibfnamefont {P.~C.}\ \bibnamefont {Humphreys}}, \bibinfo {author}
  {\bibfnamefont {R.~P.}\ \bibnamefont {Mirin}}, \bibinfo {author}
  {\bibfnamefont {J.~C.}\ \bibnamefont {Gates}}, \bibinfo {author}
  {\bibfnamefont {P.~G.~R.}\ \bibnamefont {Smith}}, \bibinfo {author}
  {\bibfnamefont {I.~A.}\ \bibnamefont {Walmsley}}, \bibinfo {author}
  {\bibfnamefont {T.}~\bibnamefont {Gerrits}}, \ and\ \bibinfo {author}
  {\bibfnamefont {S.~W.}\ \bibnamefont {Nam}},\ }\href {\doibase
  10.1364/OE.21.022657} {\bibfield  {journal} {\bibinfo  {journal} {Optics
  Express}\ }\textbf {\bibinfo {volume} {21}},\ \bibinfo {pages} {22657}
  (\bibinfo {year} {2013})}\BibitemShut {NoStop}%
\bibitem [{\citenamefont {Menicucci}(2014)}]{menicucci_fault-tolerant_2014}%
  \BibitemOpen
  \bibfield  {author} {\bibinfo {author} {\bibfnamefont {N.~C.}\ \bibnamefont
  {Menicucci}},\ }\href {\doibase 10.1103/PhysRevLett.112.120504} {\bibfield
  {journal} {\bibinfo  {journal} {Phys. Rev. Lett.}\ }\textbf {\bibinfo
  {volume} {112}},\ \bibinfo {pages} {120504} (\bibinfo {year}
  {2014})}\BibitemShut {NoStop}%
%
\bibitem {supp} See supplemental material, which includes Refs. [54-60]
\bibitem{glauber_quantum_1963}%
  \BibitemOpen
  \bibfield{author}{%
  \bibinfo {author} {\bibfnamefont{R.~J.}\ \bibnamefont{Glauber}},\ }%
  \bibfield{journal}{%
  \Doi{10.1103/PhysRev.130.2529}{\bibinfo {journal} {Phys. Rev.}}\ }%
  \textbf{\bibinfo {volume} {130}},\ \bibinfo {pages} {2529} (\bibinfo {month}
  {Jun.}\ \bibinfo {year} {1963}),\
  \bibAnnoteFile{NoStop}{glauber_quantum_1963}%
\bibitem{vogel_nonclassical_2008}%
  \BibitemOpen
  \bibfield{author}{%
  \bibinfo {author} {\bibfnamefont{W.}~\bibnamefont{Vogel}},\ }%
  \bibfield{journal}{%
  \Doi{10.1103/PhysRevLett.100.013605}{\bibinfo {journal} {Phys. Rev. Lett.}}\
  }%
  \textbf{\bibinfo {volume} {100}},\ \bibinfo {pages} {013605} (\bibinfo
  {month} {Jan.}\ \bibinfo {year} {2008}),\
  \bibAnnoteFile{NoStop}{vogel_nonclassical_2008}%
\bibitem{miranowicz_testing_2010}%
  \BibitemOpen
  \bibfield{author}{%
  \bibinfo {author} {\bibfnamefont{A.}~\bibnamefont{Miranowicz}}, \bibinfo
  {author} {\bibfnamefont{M.}~\bibnamefont{Bartkowiak}}, \bibinfo {author}
  {\bibfnamefont{X.}~\bibnamefont{Wang}}, \bibinfo {author}
  {\bibfnamefont{Y.-x.}\ \bibnamefont{Liu}},\ and\ \bibinfo {author}
  {\bibfnamefont{F.}~\bibnamefont{Nori}},\ }%
  \bibfield{journal}{%
  \Doi{10.1103/PhysRevA.82.013824}{\bibinfo {journal} {Phys. Rev. A}}\ }%
  \textbf{\bibinfo {volume} {82}},\ \bibinfo {pages} {013824} (\bibinfo {month}
  {Jul.}\ \bibinfo {year} {2010}),\
  \bibAnnoteFile{NoStop}{miranowicz_testing_2010}%
\bibitem{avenhaus_accessing_2010}%
  \BibitemOpen
  \bibfield{author}{%
  \bibinfo {author} {\bibfnamefont{M.}~\bibnamefont{Avenhaus}}, \bibinfo
  {author} {\bibfnamefont{K.}~\bibnamefont{Laiho}}, \bibinfo {author}
  {\bibfnamefont{M.~V.}\ \bibnamefont{Chekhova}},\ and\ \bibinfo {author}
  {\bibfnamefont{C.}~\bibnamefont{Silberhorn}},\ }%
  \bibfield{journal}{%
  \Doi{10.1103/PhysRevLett.104.063602}{\bibinfo {journal} {Phys. Rev. Lett.}}\
  }%
  \textbf{\bibinfo {volume} {104}},\ \bibinfo {pages} {063602} (\bibinfo
  {month} {Feb.}\ \bibinfo {year} {2010}),\
  \bibAnnoteFile{NoStop}{avenhaus_accessing_2010}%
\bibitem{sperling_uncovering_2015}%
  \BibitemOpen
  \bibfield{author}{%
  \bibinfo {author} {\bibfnamefont{J.}~\bibnamefont{Sperling}}, \bibinfo
  {author} {\bibfnamefont{M.}~\bibnamefont{Bohmann}}, \bibinfo {author}
  {\bibfnamefont{W.}~\bibnamefont{Vogel}}, \bibinfo {author}
  {\bibfnamefont{G.}~\bibnamefont{Harder}}, \bibinfo {author}
  {\bibfnamefont{B.}~\bibnamefont{Brecht}}, \bibinfo {author}
  {\bibfnamefont{V.}~\bibnamefont{Ansari}},\ and\ \bibinfo {author}
  {\bibfnamefont{C.}~\bibnamefont{Silberhorn}},\ }%
  \bibfield{journal}{%
  \Doi{10.1103/PhysRevLett.115.023601}{\bibinfo {journal} {Phys. Rev. Lett.}}\
  }%
  \textbf{\bibinfo {volume} {115}},\ \bibinfo {pages} {023601} (\bibinfo
  {month} {Jul.}\ \bibinfo {year} {2015}),\
  \bibAnnoteFile{NoStop}{sperling_uncovering_2015}%
\bibitem{feito_measuring_2009}%
  \BibitemOpen
  \bibfield{author}{%
  \bibinfo {author} {\bibfnamefont{A.}~\bibnamefont{Feito}}, \bibinfo {author}
  {\bibfnamefont{J.~S.}\ \bibnamefont{Lundeen}}, \bibinfo {author}
  {\bibfnamefont{H.}~\bibnamefont{Coldenstrodt-Ronge}}, \bibinfo {author}
  {\bibfnamefont{J.}~\bibnamefont{Eisert}}, \bibinfo {author}
  {\bibfnamefont{M.~B.}\ \bibnamefont{Plenio}},\ and\ \bibinfo {author}
  {\bibfnamefont{I.~A.}\ \bibnamefont{Walmsley}},\ }%
  \bibfield{journal}{%
  \Doi{10.1088/1367-2630/11/9/093038}{\bibinfo {journal} {New J. Phys.}}\ }%
  \textbf{\bibinfo {volume} {11}},\ \bibinfo {pages} {093038} (\bibinfo {month}
  {Sep.}\ \bibinfo {year} {2009}),\ ISSN \bibinfo {issn} {1367-2630},\
  \bibAnnoteFile{NoStop}{feito_measuring_2009}%
\bibitem{levine_algorithm_2012}%
  \BibitemOpen
  \bibfield{author}{%
  \bibinfo {author} {\bibfnamefont{Z.~H.}\ \bibnamefont{Levine}}, \bibinfo
  {author} {\bibfnamefont{T.}~\bibnamefont{Gerrits}}, \bibinfo {author}
  {\bibfnamefont{A.~L.}\ \bibnamefont{Migdall}}, \bibinfo {author}
  {\bibfnamefont{D.~V.}\ \bibnamefont{Samarov}}, \bibinfo {author}
  {\bibfnamefont{B.}~\bibnamefont{Calkins}}, \bibinfo {author}
  {\bibfnamefont{A.~E.}\ \bibnamefont{Lita}},\ and\ \bibinfo {author}
  {\bibfnamefont{S.~W.}\ \bibnamefont{Nam}},\ }%
  \bibfield{journal}{%
  \Doi{10.1364/JOSAB.29.002066}{\bibinfo {journal} {J. Opt. Soc. Am. B}}\ }%
  \textbf{\bibinfo {volume} {29}},\ \bibinfo {pages} {2066} (\bibinfo {month}
  {Aug.}\ \bibinfo {year} {2012}),\
  \bibAnnoteFile{NoStop}{levine_algorithm_2012}%
\end{thebibliography}

%
\end{document}